\documentclass[format=sigconf]{acmart}
\renewcommand\footnotetextcopyrightpermission[1]{}
\usepackage{graphicx}
\usepackage{subfig}
\usepackage{enumerate}
\usepackage{multirow}
\usepackage{booktabs}
\usepackage{amsmath}
\usepackage{balance}
\usepackage{epstopdf}
\usepackage{color}

\usepackage{amssymb}
\usepackage{bm}
\begin{document}


\title{SPQE: Structure-and-Perception-Based Quality Evaluation for Image Super-Resolution}


%
%
\author{Keke Zhang}
\email{201110007@fzu.edu.cn}
\affiliation{%
 \institution{Fuzhou University}
 \country{ }}
%

\author{Tiesong Zhao}
\email{t.zhao@fzu.edu.cn}
\affiliation{%
 \institution{Fuzhou University}
 \country{ }}

\author{Weiling Chen}
\email{weiling.chen@fzu.edu.cn}
\affiliation{%
 \institution{Fuzhou University}
 \country{ }}

\author{Yuzhen Niu}
\email{yuzhenniu@gmail.com}
\affiliation{%
 \institution{Fuzhou University}
 \country{ }}

\author{Jinsong Hu}
\email{jinsong.hu@fzu.edu.cn}
\affiliation{%
 \institution{Fuzhou University}
 \country{ }}

%
%
%
%



\begin{abstract}
The image Super-Resolution (SR) technique has greatly improved the visual quality of images by enhancing their resolutions. It also calls for an efficient SR Image Quality Assessment (SR-IQA) to evaluate those algorithms or their generated images. In this paper, we focus on the SR-IQA under deep learning and propose a Structure-and-Perception-based Quality Evaluation (SPQE). In emerging deep-learning-based SR, a generated high-quality, visually pleasing image may have different structures from its corresponding low-quality image. In such case, how to balance the quality scores between no-reference perceptual quality and referenced structural similarity is a critical issue. To help ease this problem, we give a theoretical analysis on this tradeoff and further calculate adaptive weights for the two types of quality scores. We also propose two deep-learning-based regressors to model the no-reference and referenced scores. By combining the quality scores and their weights, we propose a unified SPQE metric for SR-IQA. Experimental results demonstrate that the proposed method outperforms the state-of-the-arts in different datasets.
\end{abstract}



\maketitle

\section{Introduction}

\begin{figure}[t]
	\centering
	\subfloat [Original image]{
		\includegraphics[width=1.9cm]{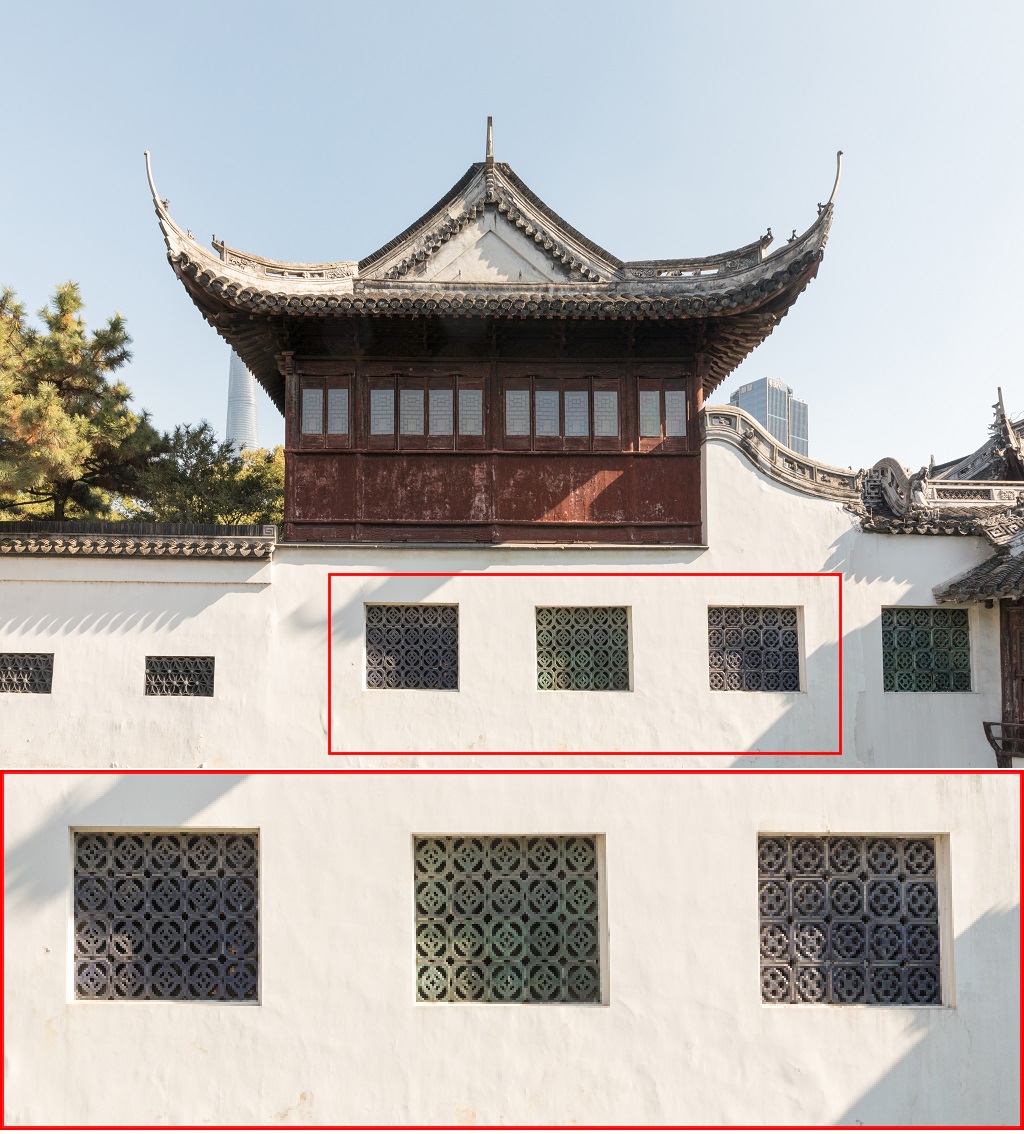} }\hspace{-1mm}
	\subfloat [0.8363/0.7395]{
		\includegraphics[width=1.9cm]{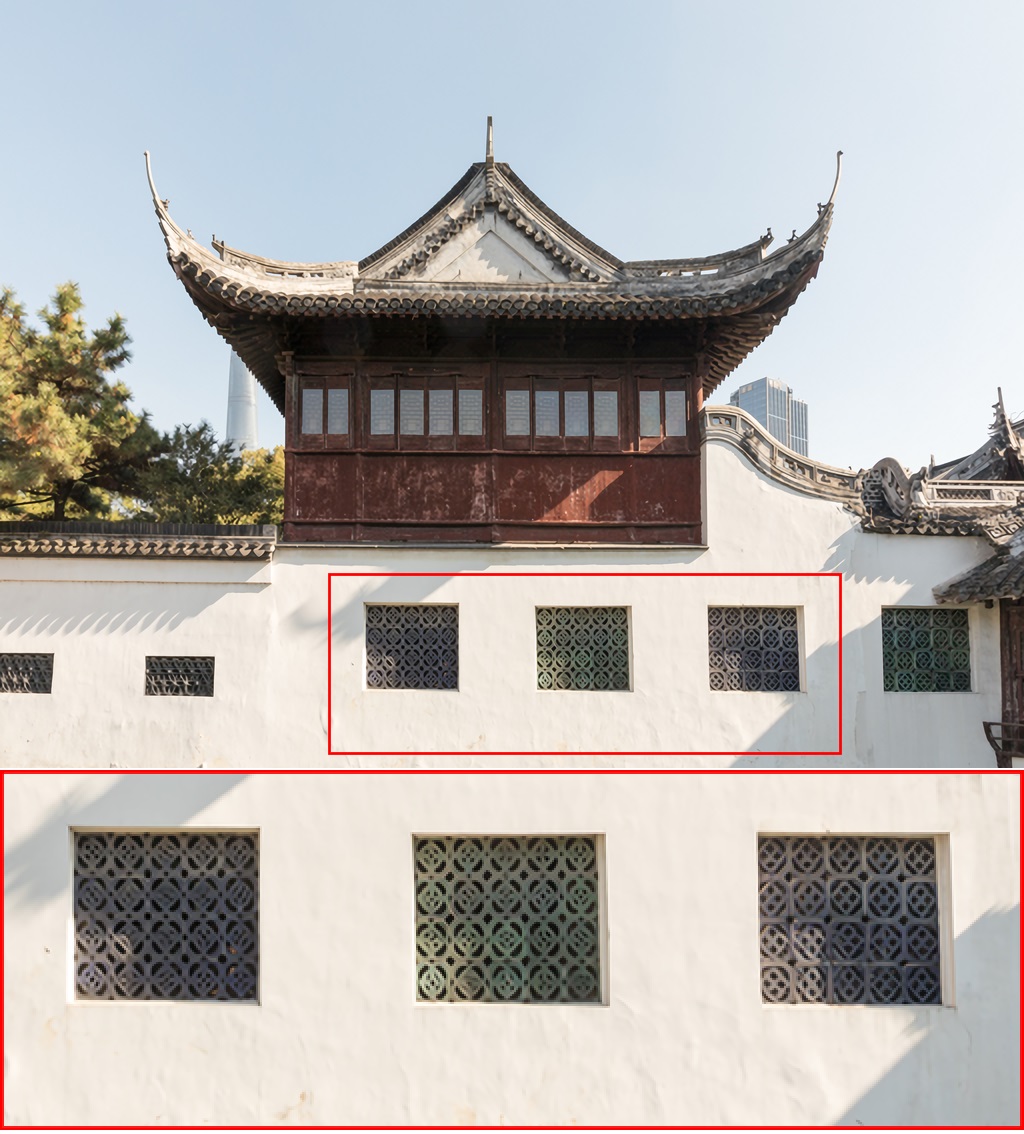} }\hspace{-1mm}
	\subfloat [0.8665/0.4847]{
		\includegraphics[width=1.9cm]{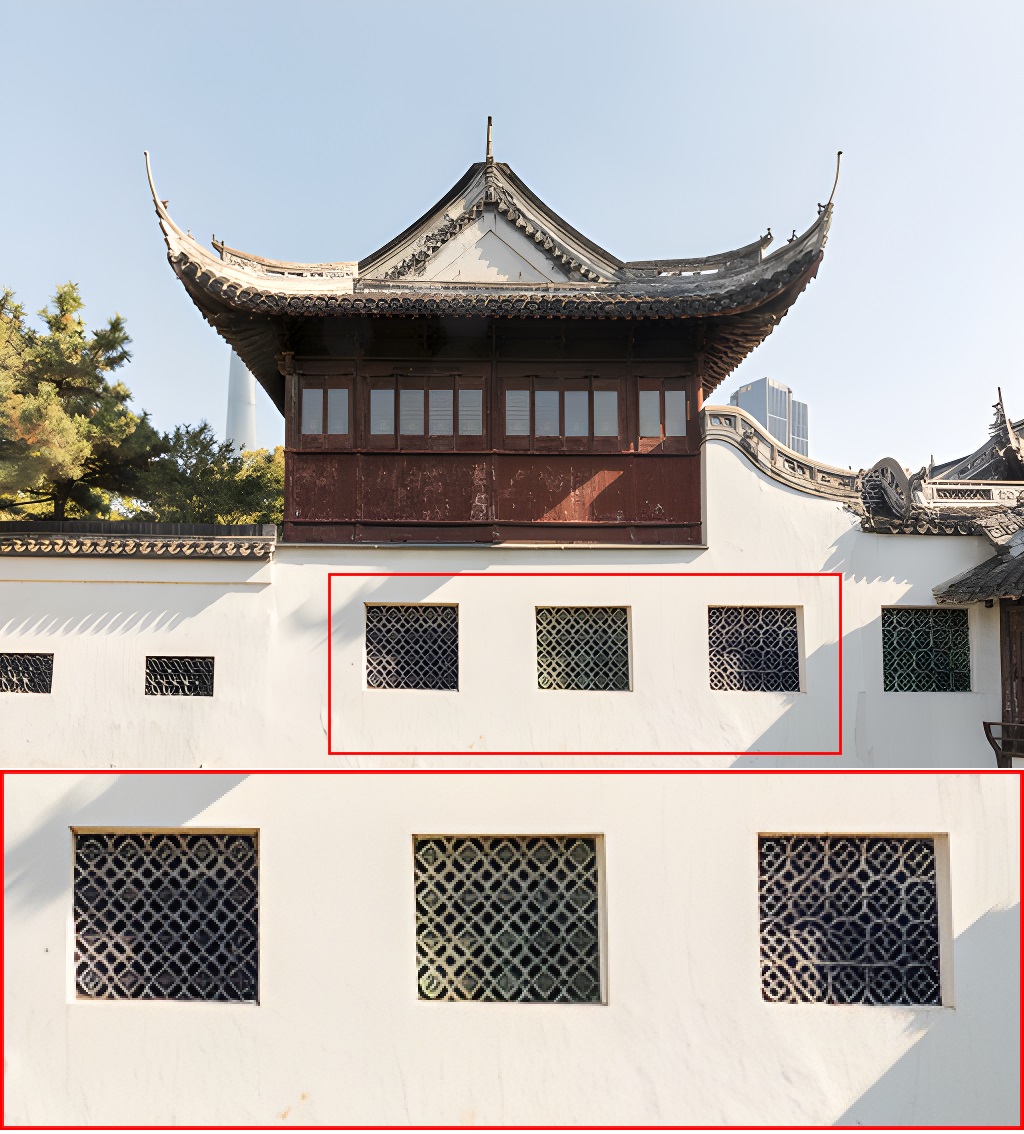} }\hspace{-1mm}
	\subfloat [0.8200/0.7335]{
		\includegraphics[width=1.9cm]{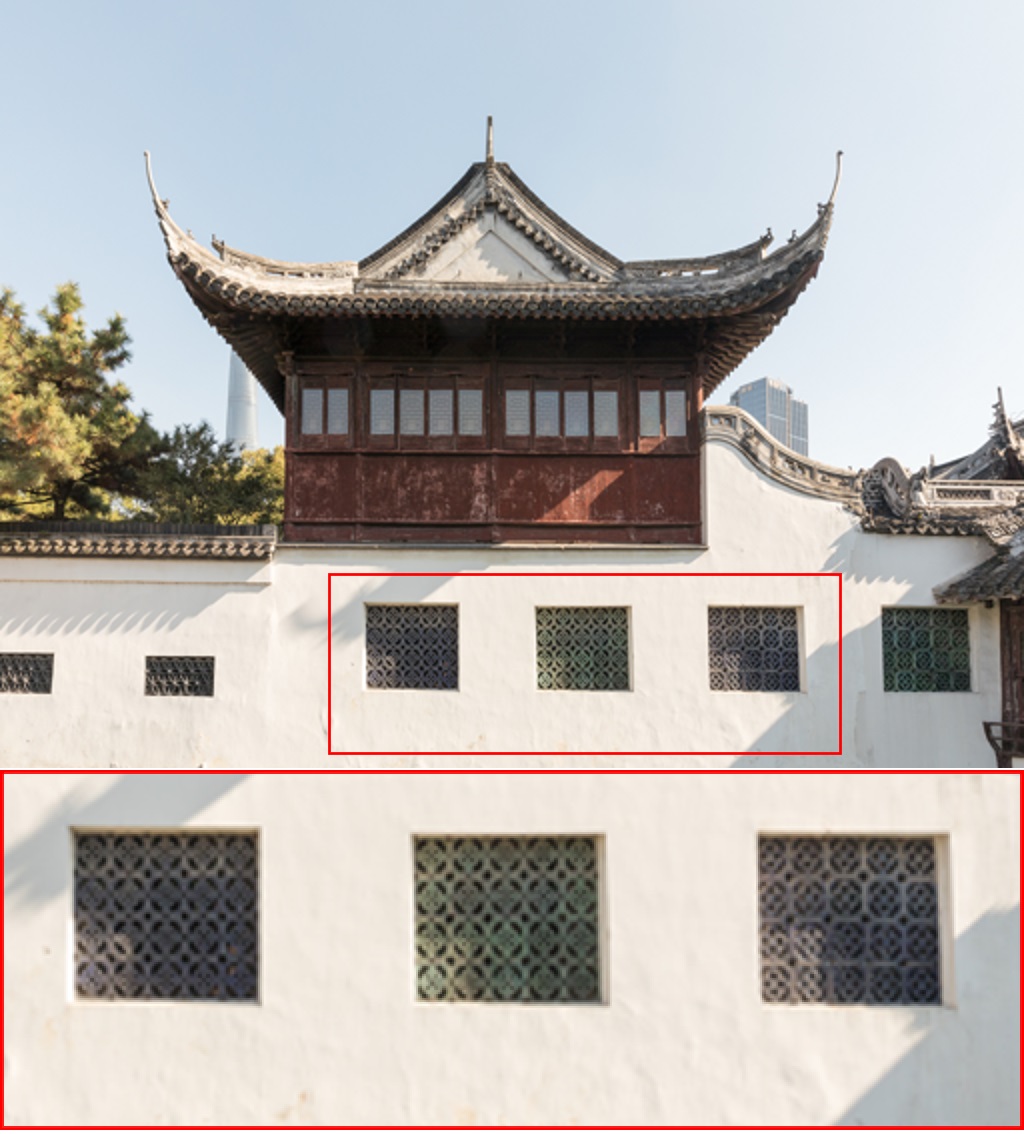} } \\
	
	\subfloat [Original image]{
		\includegraphics[width=1.9cm]{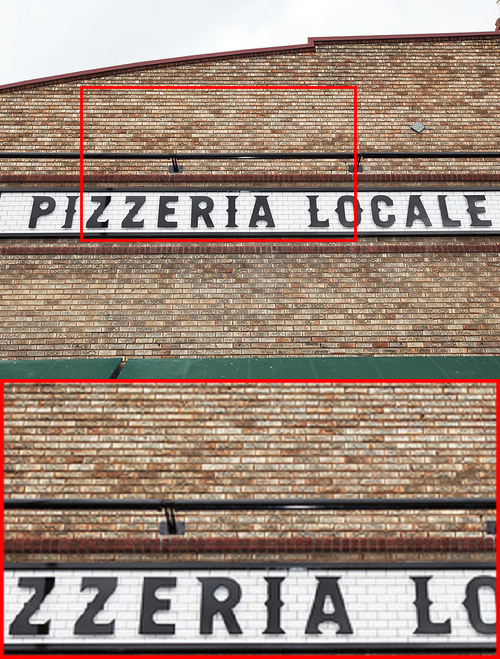} }\hspace{-1mm}
	\subfloat [0.8469/0.6990]{
		\includegraphics[width=1.9cm]{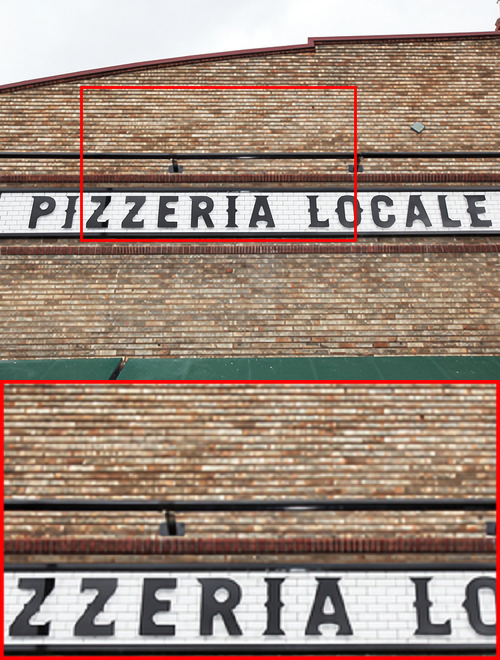} }\hspace{-1mm}
	\subfloat [0.8692/0.5967]{
		\includegraphics[width=1.9cm]{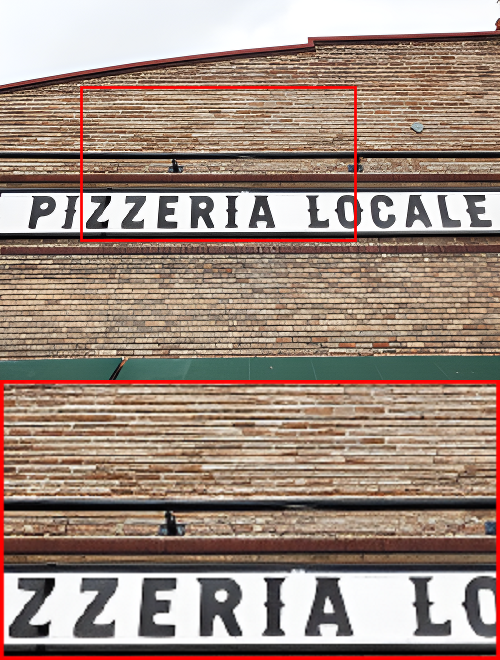} }\hspace{-1mm}
	\subfloat [0.5960/0.7435]{
		\includegraphics[width=1.9cm]{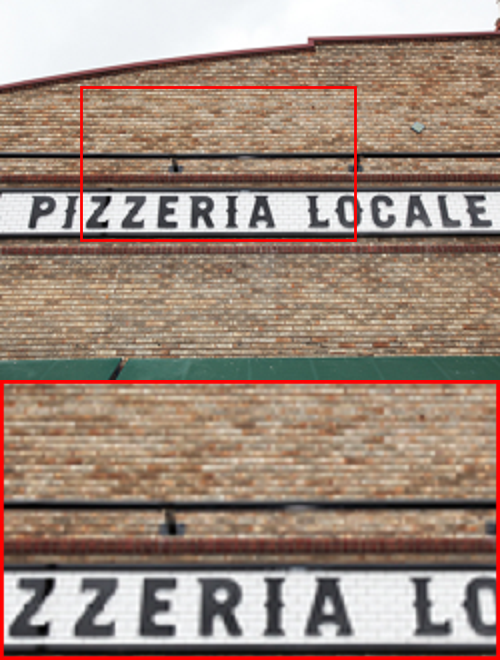} } \\
	\vspace{-0.3cm}
	\caption{Examples of SR images. The first column is the original HR images, while the others are the SR images generated by RCAN \cite{RCAN}, Real-ESRGAN \cite{Real-ESRGAN} and Bicubic with scaling factor 2, respectively. The quality scores of SR images are separately calculated by DISQ \cite{SISAR} and DeepSRQ \cite{DeepSRQ}. How to effectively assess the qualities of SR images remains a challenging task.}
	\label{fig:1}
	\vspace{-0.8mm}
\end{figure}

Image Super-Resolution (SR) refers to reconstruct High-Resolution (HR) images from the corresponding Low-Resolution (LR) images. Recent years have witnessed remarkable advances in image SR, with a wide range of applications including video surveillance, medical imaging, satellite imaging, \emph{etc}. In image SR, the quality of SR images vary significantly when different SR algorithms are employed. Thus, an Image Quality Assessment (IQA) is imperative to evaluate the quality degradations of SR images and guide the further developments of SR methods.

Existing IQA methods can be classified into subjective and objective methods. Between them, subjective evaluation is the most credible, since humans are the ultimate receivers of the majority of images. However, it is also laborious, expensive, and cannot be integrated into practical SR optimization systems. As a result, the subjective evaluation is usually employed to construct datasets and serves as a golden metric in comparing objective IQA methods. It is essential to develop objective SR-IQA metrics that are consistent with the subjective evaluation. 

According to the availability of unimpaired reference images, objective IQA methods can be divided into Full-Reference (FR), Reduced-Reference (RR) and No-Reference (NR) methods. Despite the advances in other IQA applications, existing conventional IQA methods still fall short in evaluating the quality of SR images. Existing SR algorithms often lead to hybrid impairments, including ringing, blurring, and aliasing, \emph{etc}, which are not well handled by the existing universal IQA methods \cite{CVIU}, \cite{SISAR}. 

There have been some SR-IQA methods in recent years, including handcrafted feature-based methods \cite{WIND, HYQM, SFSN} and learning-based methods \cite{CVIU,CNNSR,DeepSRQ,SISAR}. Those methods were proved to achieve good performances to evaluate the non-deep SR algorithms. However, in the emerging deep-learning-based SR, a generated visually pleasing image may have different structures from those of its LR reference, which might lead to disagreements in existing SR-IQA approaches. As shown in Fig. \ref{fig:1}, the SR images generated from Real-ESRGAN (Fig. \ref{fig:1} (c) (g)) have apparently different structures to the original images, but these images obtain better DISQ scores. In addition, the SR images generated from Bicubic (Fig. \ref{fig:1} (d) (h)) have severe blurring artifacts, but these images obtain better DeepSRQ scores. This may be because the DISQ and DeepSRQ methods evaluate the quality of SR images only from the perspective of referenced and no-reference, respectively, and do not consider the tradeoff between them. Therefore, it is imperative to design a tradeoff mechanism to balance the no-reference and referenced quality scores of SR-IQA.   

Notably, the difference between the structural similarity and perceptual quality explored in this work is clarified as follows. We define the \emph{structural similarity} as the visual similarity between SR image and the corresponding LR reference. Instead, the \emph{perceptual quality} refers only to the visual quality of SR image, regardless of its similarity to LR reference. As evidenced in \cite{tradeoffcvpr}, image restoration algorithms should be evaluated by a pair of NR and FR metrics (please refer to \cite{tradeoffcvpr} for details). However, this method has two drawbacks: (i) it is cumbersome to employ NR and FR metrics separately; (ii) human prefers a single score to measure and compare image quality. Inspired by this, we aim at overcoming the following two challenges in the scenario of SR-IQA: (i) how to effectively measure the no-reference and referenced quality scores simultaneously; (ii) how to balance the two types of quality scores.   

In summary, the main contributions of this paper are as follows:
\begin{enumerate}[1)]
	\item We give a theoretically analysis on how to balance the quality scores between the no-reference perceptual quality and the referenced structural similarity of SR images. To the best of our knowledge, this is the first attempt to consider this tradeoff in SR-IQA.
	
	\item We design two deep-learning-based regressors to model the no-reference and referenced scores for SR-IQA. The two regressors leverage multi-scale and saliency information to evaluate the structural similarity and perceptual quality of SR images, respectively.
	
	\item We develop a unified, Structure-and-Perception-based Quality Evaluation (SPQE) with the tradeoff and regressors. Massive experiments demonstrate the superiority of the proposed method against the state-of-the-arts in four benchmark datasets.  
\end{enumerate}

\section{RELATED WORK}
\textbf{SR methods}. There are three categories of SR methods. \emph{Interpolation-based methods} aim at generating SR images by filling missing pixels on LR images. The classic algorithms include nearest-neighbor, bilinear, and bicubic. \emph{Reconstruction-based methods} are built on models of prior domain knowledge. Papyan \emph{et al.} \cite{srchonggou2016tip} designed a local low dimensional prior on scale-patches for image SR. Ren \emph{et al.} \cite{srchonggou2017tip} proposed a high-dimensional non-local total variation prior for image SR. \emph{Learning-based methods} essentially learn the mapping from LR images to SR images. Early learning-based SR methods are mainly based on sparse coding \cite{xishubianma2010} and neighbor embedding \cite{qianru2004}. Recent years have witnessed numerous deep-learning-based SR methods, including Convolutional Neural Network (CNN) based methods \cite{SRCNN,RCAN, SAN, SR2021TMM} and Generative Adversarial Network (GAN) based methods \cite{SRGAN,ESRGAN,Real-ESRGAN,RankSRGAN}. Dong \emph{et al.} \cite{SRCNN} proposed SRCNN to image SR by an end-to-end learning mode. Ledig \emph{et al.} \cite{SRGAN} proposed the first GAN-based SR, named SRGAN. Subsequently, GAN-based algorithms have emerged rapidly, such as ESRGAN \cite{ESRGAN}, Real-ESRGAN \cite{Real-ESRGAN}, and RankSRGAN \cite{RankSRGAN}. 

\textbf{Two-step IQA methods}. The \emph{two-step quality prediction} concept is defined in \cite{tradeoff2019tip}, which refers to the IQA methods that combines no-reference and referenced quality measures. The typical process of those methods is: (i) design two types of IQA methods separately; (ii) design a fixed strategy to fuse the two types of quality scores. Yu \emph{et al.} \cite{tradeoff2019tip} utilized an NR-IQA to determine the degraded quality of original image first, then adopted the obtained information in the next FR-IQA step, which calculated the perceptual difference between original and compressed images. Subsequently, a weighted product method was designed to combine the quality scores. Yeganeh \emph{et al.} \cite{tradeoff2013tip} developed an NR metric to compute statistical naturalness of tone-mapped Low Dynamic Range (LDR) image, and an FR metric to measure structural fidelity between High Dynamic Range (HDR) and tone-mapped LDR images. Then a three-parameter function was designed to fuse the above two measures. Similar to \cite{tradeoff2013tip}, Zhou \emph{et al.} \cite{SFSN} designed an NR metric to evaluate statistical naturalness of SR image, and an FR metric to measure structural fidelity between original HR images and SR images. A linear combination is adopted to fuse the two quality scores. 

Notably, our SPQE metric is completely different from the two-step IQA methods. First, the SPQE can optimize the no-reference (perception) and referenced (structure) quality score regressors simultaneously. Second, the adaptive weight regressor can calculate adaptive weights to balance the two types of quality scores.

\textbf{SR-IQA methods}. Existing SR-IQA methods can be divided into handcrafted feature-based \cite{WIND, CVIU, HYQM, SFSN} and deep-learning-based \cite{CNNSR, DeepSRQ, SISAR} methods. \emph{Handcrafted feature-based methods} are devoted to capturing features for quality evaluation manually. Yeganeh \emph{et al.} \cite{WIND} integrated the frequency energy falloff, dominant orientation, and spatial continuity from both LR and SR images to measure the quality of integer-interpolated images. Ma \emph{et al.} \cite{CVIU} extracted three types of statistical features from both frequency and spatial domains to evaluate SR images without the knowledge of reference images. Chen \emph{et al.} \cite{HYQM} proposed a hybrid quality metric for non-integer image interpolation that extracted features from both reduced- and no-reference scenes. \emph{Deep-learning-based methods} aim at automatically learning the mapping from image to quality. Fang \emph{et al.} \cite{CNNSR} proposed a blind quality evaluation by a simple CNN, which only consists of eight layers. Zhou \emph{et al.} \cite{DeepSRQ} designed an NR-IQA using a two-stream CNN, which directly adopts the extracted structure and texture images from SR images as inputs. Zhao \emph{et al.} \cite{SISAR} designed a Deep Image SR Quality (DISQ) model by employing a two-stream CNN. The inputs of the DISQ model are SR and LR images, respectively. 

Despite recent advances, the current SR-IQA methods are still limited in evaluating the quality of SR images. Those methods do not consider the tradeoff between no-reference and referenced scores when evaluating the quality of SR images. Motivated by this, we give a theoretical analysis on this tradeoff, and then propose a unified, end-to-end SPQE metric for SR-IQA.

\section{PROBLEM STATEMENT}
SR images are oriented to both no-reference and referenced scenes. For example, SR image is reference-free in practical use, while has reference image in SR task. As discussed above, the emerging deep-learning-based SR methods may generate visually pleasing HR images but with different structure from its LR reference. As a result, how to effectively balance the quality scores between the no-reference perceptual quality and the referenced structural similarity is crucial in SR-IQA.

\begin{figure}[t]
	\centering
	\subfloat [Original]{
		\includegraphics[width=1.9cm]{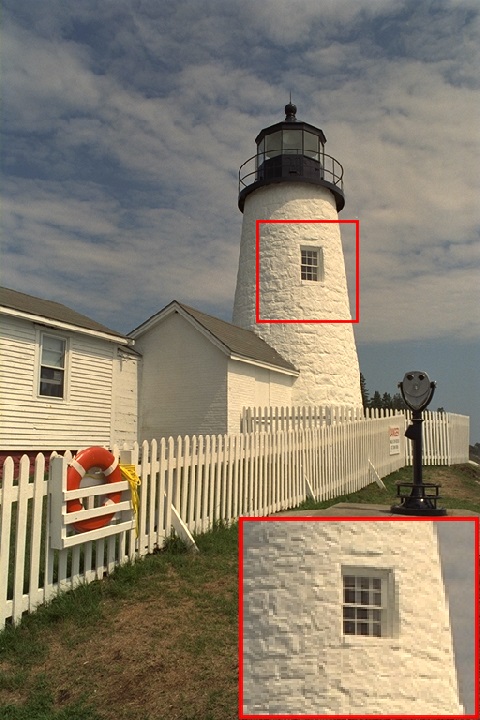} }\hspace{-1mm}
	\subfloat [$I_{A}$/0.1370]{
		\includegraphics[width=1.9cm]{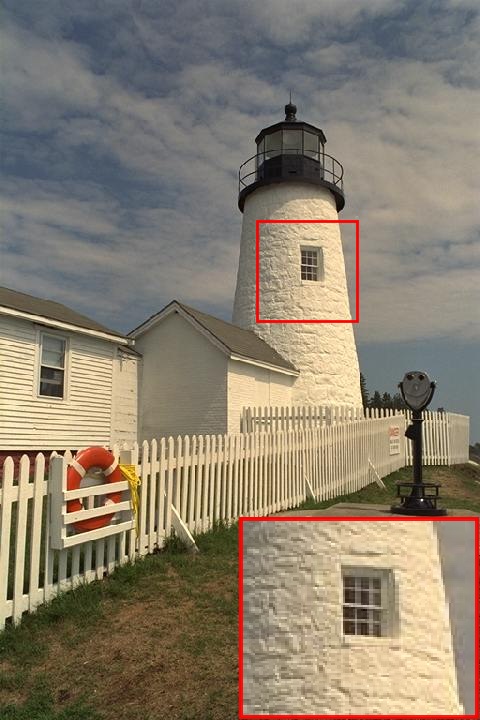} }\hspace{-1mm}
	\subfloat [$I_{B}$/0.3117]{
		\includegraphics[width=1.9cm]{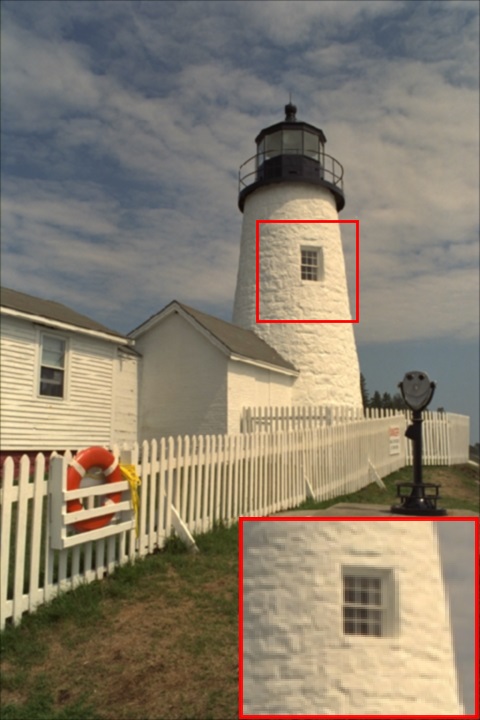} }\hspace{-1mm}
	\subfloat [$I_{C}$/0.8135]{
		\includegraphics[width=1.9cm]{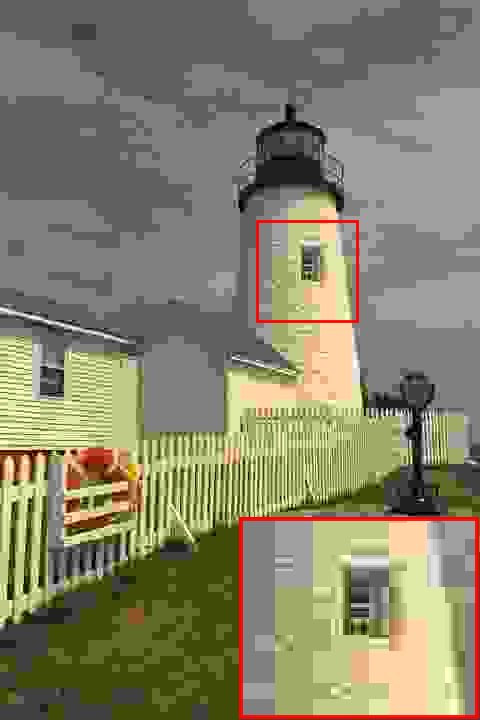} } \\
	\vspace{-2mm}
	\subfloat [Original/$I_{A}$]{
		\includegraphics[width=2.6cm]{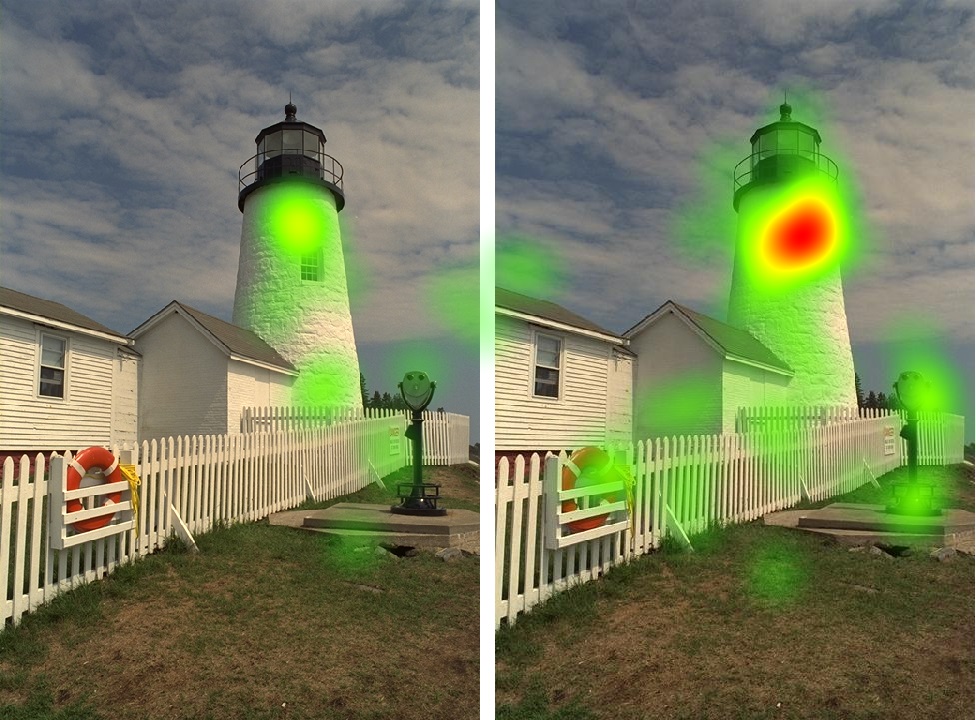} }\hspace{-1mm}
	\subfloat[Original/$I_{B}$]{
		\includegraphics[width=2.6cm]{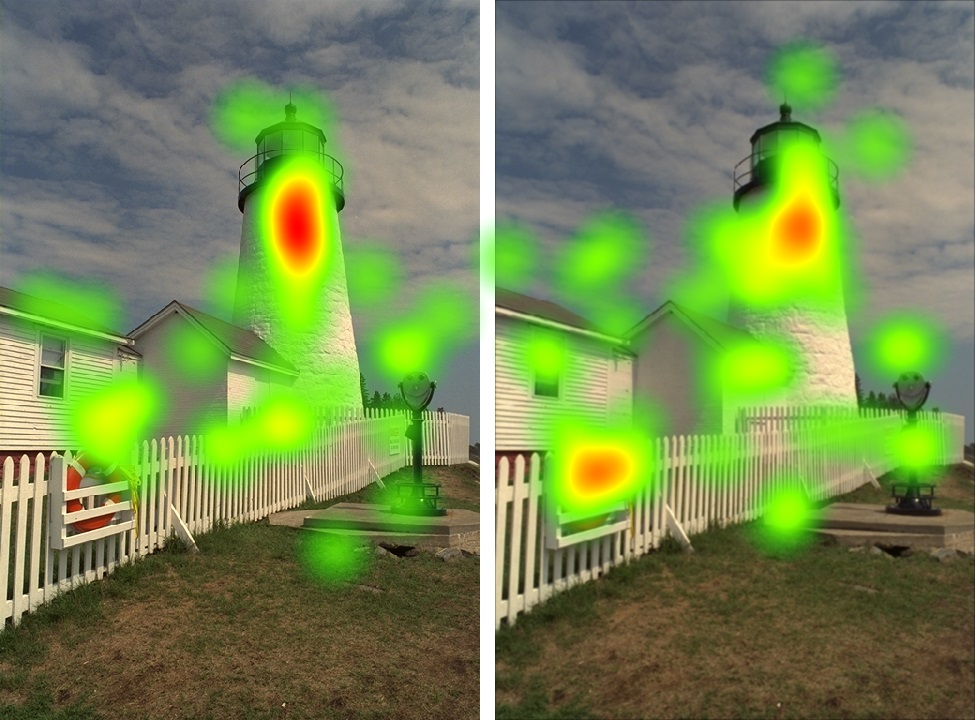} }\hspace{-1mm}
	\subfloat[Original/$I_{C}$]{
		\includegraphics[width=2.6cm]{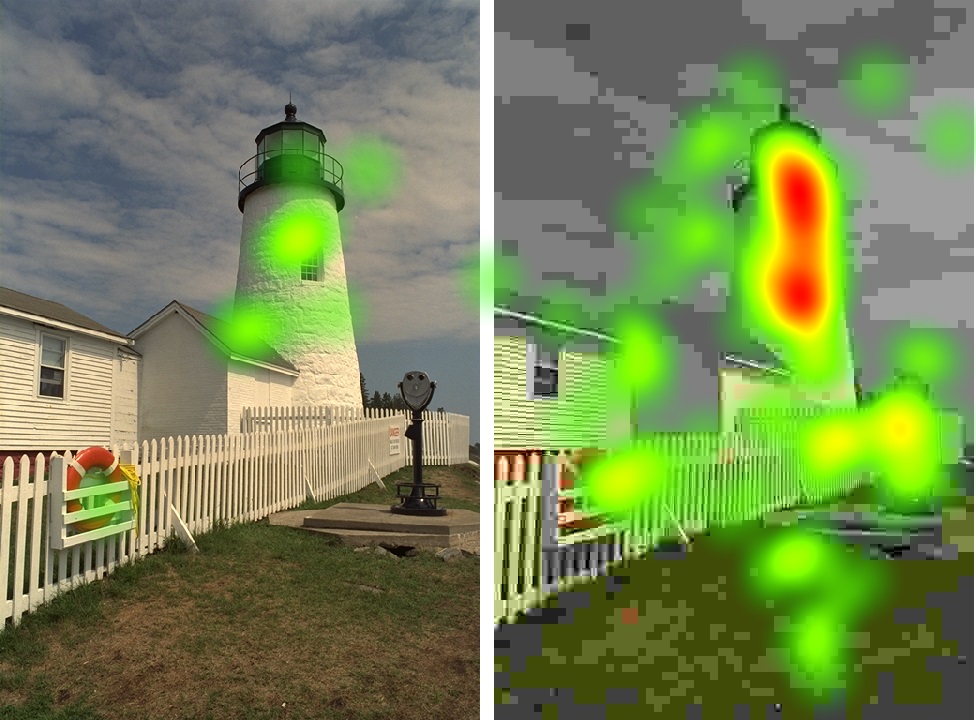} }\\
	\vspace{-0.3cm}
	\caption{Examples of original and distorted image and their corresponding visual heat maps in quality evaluation.}
	\label{fig:add2}
\end{figure}

To explore this tradeoff mechanism, we resort to the properties of Human Visual System (HVS). HVS is a highly adaptive system that adopts multiple strategies when determining quality. \cite{MAD} reveals two useful HVS properties related to the images with near-threshold distortions (i) and clearly visible distortions (ii). We empiracally extend the (ii) to a reasonable one (iii), which has been confirmed in the eye tracker experiment (See Fig. \ref{fig:add2}). The three HVS characteristics are:

\textbf{(i)} When judging the quality of an image with near-threshold distortions, the HVS attempts to locate differences using point-by-point comparisons with the reference image.

\textbf{(ii)} When judging the quality of an image with clearly visible distortions, the HVS focuses on recognizing the image content than difference detection. 

\textbf{(iii)} When judging the quality of an image with imperceptible distortions, the HVS tends to rely much more on image content and personal preference than difference detection. 
\begin{table}[t]
	\centering
	\caption{Correlation between image artifacts and distortions}
	\label{table:add1}
	\setlength{\tabcolsep}{0.8mm}
	\begin{tabular}{@{}c|c|c|ccccc@{}}
		\toprule
		Dataset&Artifacts&DMOS&SSIM&MS-SSIM&IFC&FSIM&GMSD\\
		\midrule
		\multirow{2}{*}{LIVE}&jpeg&0.974&0.932&0.983&0.962&0.989&0.973\\
		&blur&0.956&0.861&0.953&0.982&0.963&0.963\\
		\midrule
		\multirow{2}{*}{CSIQ}&jpeg&0.932&0.908&0.960&0.952&0.973&0.967\\
		&blur&0.906&0.870&0.926&0.963&0.935&0.944\\
		\bottomrule
	\end{tabular}
\end{table}

Intuitively, as shown in Fig. \ref{fig:add2} (images are from the LIVE dataset \cite{LIVE}), $I_A$, $I_B$, and $I_C$ contain imperceptible, near-threshold, and clearly visible distortions, respectively. The values of images represent Differential Mean Opinion Scores (DMOS). The second row represents the average visual heat maps (five participants) from eye tracker equipment (Tobii Pro Spectrum 150) when evaluating the three images. From Fig. \ref{fig:add2}, the human eye fixations in $I_A$ and $I_{C}$ significantly exceed those in the original images. Instead, the human eye fixations in $I_{B}$ is similar to those of the original image. These findings confirm the three HVS properties. Thus, when evaluating the three images, their weights between the no-reference perceptual quality and referenced structural similarity scores should be different. Denote the perception weights as $W_{p,A}$, $W_{p,B}$, and $W_{p,C}$, respectively. From the three HVS characteristics, they should satisfy $W_{p,A} > W_{p,B}$ and $W_{p,C} > W_{p,B}$.

The above HVS properties are defined in a referenced scene, while the distortions are hard to evaluate in no-reference scenarios. Hence, we introduce the visual artifacts as an alternate. Considering that 1) image artifacts are crucial to quality assessment in IQA community; 2) the three HVS characteristics has been established based on distortions, we will investigate the correlation between image artifacts and distortions. Specifically, the images in \emph{jpeg} and \emph{blur} folders of LIVE \cite{LIVE} and CSIQ \cite{MAD} datasets are adopted, since the artifacts of those images are common and the corresponding measures are mature. We adopt two widely accepted task-specific NR measures \cite{jpeg} (for \emph{jpeg}) and \cite{blur} (for \emph{blur}) to capture image artifacts, and five representative FR IQA algorithms to evaluate image distortions, including SSIM \cite{SSIM}, MS-SSIM \cite{MS-SSIM}, IFC \cite{IFC}, FSIM \cite{FSIM}, and GMSD \cite{GMSD}. The Spearman Rankorder Correlation Coefficient (SRCC) criterion is employed to evaluate the correlation between image artifacts and distortions, see Table \ref{table:add1}.

From Table \ref{table:add1}, we have the following findings. First, the correlation between artifacts measures and DMOS is high. It indicates that image artifacts are well captured by the employed algorithms. Second, the overall correlation between image artifacts and distortions is preferable. Obviously, the correlation between the well captured artifacts and distortions is quite high, such as the jpeg \emph{vs.} FSIM, and the blur \emph{vs.} IFC in LIVE. As a result, the image artifacts are highly correlated with distortions. The distortion-related HVS properties can be extended to artifacts-related. 

To embed the artifacts-related HVS properties in our SPQE metric, we add a weight to the perceptual quality term. Notably, the \emph{weight} is artifacts-aware, and can be adaptively adjusted for different images. Denote the \emph{final} perceptual quality of an image as $P_{p}$, for an image $I_s$, the $P_{p}$ can be computed by:
\begin{equation}
P_p(I_s)=f_w(A_s)\cdot f_p(I_s), s.t. \ A_s \gets f_a(I_s),
\label{func:add4}
\end{equation} 
where $f_{w}$, $f_{p}$, and $f_{a}$ represent the adaptive perception weight regressor, perception score regressor, and artifacts extraction model, respectively. $A_{s}$ denotes the artifacts-related feature. 

In this paper, the structure and perception serve as complementary one to the other in evaluating the quality of SR images. Thus the structure weight $W_{s}$ varies with perception weight $W_{p}$, and can be computed as $1-W_p$. By combining the structure and perception quality scores and their weights, we propose a unified, end-to-end SPQE metric for SR-IQA. The SPQE can measure the perception (no-reference) and structure (referenced) scores of SR images simultaneously, and adaptively balance the two types of scores:
\begin{equation}
\begin{aligned}
S_{spqe}=W_{p}\cdot S_{p}+(1-W_p)\cdot S_{s},
\end{aligned}
\label{func:10}
\end{equation}
where the $W_{p}$ and $S_{p}$ correspond to the $f_{w}(\cdot)$ and $f_{p}(\cdot)$ in Eq. (\ref{func:add4}), respectively. $S_{s}$ denotes the structure score. The purpose of SPQE metric is to minimize the difference between the predicted quality score and the ground truth $S_{gt,i}$ of the $i$th image, that is
\begin{equation}
\begin{aligned}
{\rm min} \dfrac{1}{N}\sum_{i=1}^N \Vert [W_{p,i}\cdot S_{p,i}+(1-W_{p,i})\cdot S_{s,i}]-S_{gt,i} \Vert_1.
\end{aligned}
\label{func:add6}
\end{equation}
Subsequently, the $l$-1 norm in Eq. (\ref{func:add6}) is adopted as the loss function in the training stage of our SPQE metric.
\section{PROPOSED METHOD}
\begin{figure}[t]
	\centering
	\includegraphics[width=8.8cm]{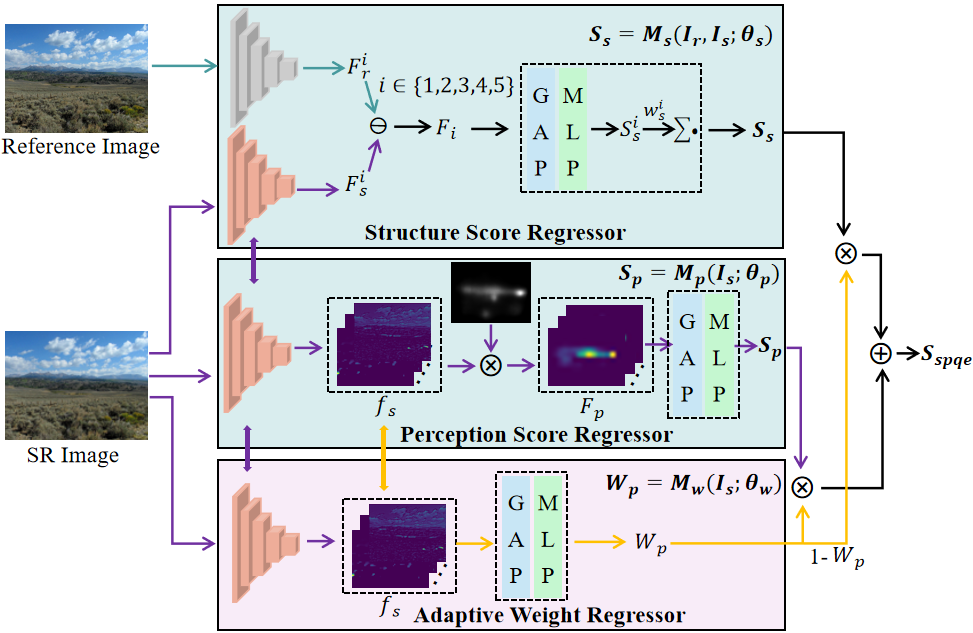}
	\caption{The framework of our unified SPQE metric. The structure score and perception score regressors are designed to model the referenced structural similarity and no-reference perceptual quality scores, respectively. The adaptive weight regressor is developed to calculate adaptive weights to balance the two types of quality scores.}  
	\label{fig:2}
\end{figure}
\subsection{Framework Overview}
Our goal is to design a unified, end-to-end SPQE metric for SR-IQA (see Fig. \ref{fig:2}). The SPQE model consists of two quality score regressors and an adaptive weight regressor. The former is to obtain the perception (no-reference) and structure (referenced) scores, and the latter is to calculate the adaptive weights for the two types of quality scores.

\subsection{Quality Score Regressors}
\textbf{Structure score regressor}. It aims to calculate the structure score $S_s$ between SR image $I_s$ and its reference image $I_r$, and can be described as:
\begin{equation}
S_s=M_s(I_r,I_s;\theta_s),	
\label{func:1}
\end{equation}
where $M_s$ and $\theta_{s}$ represent the structure score regressor and its parameters. Considering that 1) structural degradations may occur in local or global regions of SR images; 2) the quality evaluation of an image is closely related to the viewing conditions, such as display resolution and viewing distance \cite{MS-SSIM}, we extract multi-scale information for quality assessment. To this end, we develop a multi-scale quality regressor $M_s$ using CNN. There are two reasons why we use CNN as the base of our SPQE metric: 1) to leverage its strong feature learning ability; 2) to suffice the end-to-end requirements of the SPQE. To be specific, we adopt VGG16 \cite{VGG16} as the backbone of all networks, and the weights in backbone network are initialized from its ImageNet \cite{Imagenet} pre-trained model. In practice, the $M_s$ includes three parts: feature extraction, feature fusion, and quality score regression (five scores of five scales).

With respect to feature extraction of $M_s$, we extract features ($F_r$ and $F_s$) from five stages of $M_s$ for $I_s$ and $I_r$:
\begin{equation}
\begin{aligned}
& F_r^i \gets M_s(I_r;\theta_s),i\in \{1,2,3,4,5\},\\
& F_s^i \gets M_s(I_s;\theta_s),i\in \{1,2,3,4,5\}.
\end{aligned}
\label{func:2}
\end{equation}
Then, to calculate the difference between $F_r^i$ and $F_s^i$, we follow \cite{bosse} and employ a subtraction operation:
\begin{equation}
\begin{aligned}
F_i \gets F_r^i - F_s^i,i\in \{1,2,3,4,5\},
\end{aligned}
\label{func:3}
\end{equation}  
where $F_i$ indicates the fused feature of $M_s$.

\begin{figure}[t]
	\centering
	\subfloat []{
		\includegraphics[height=4.85cm]{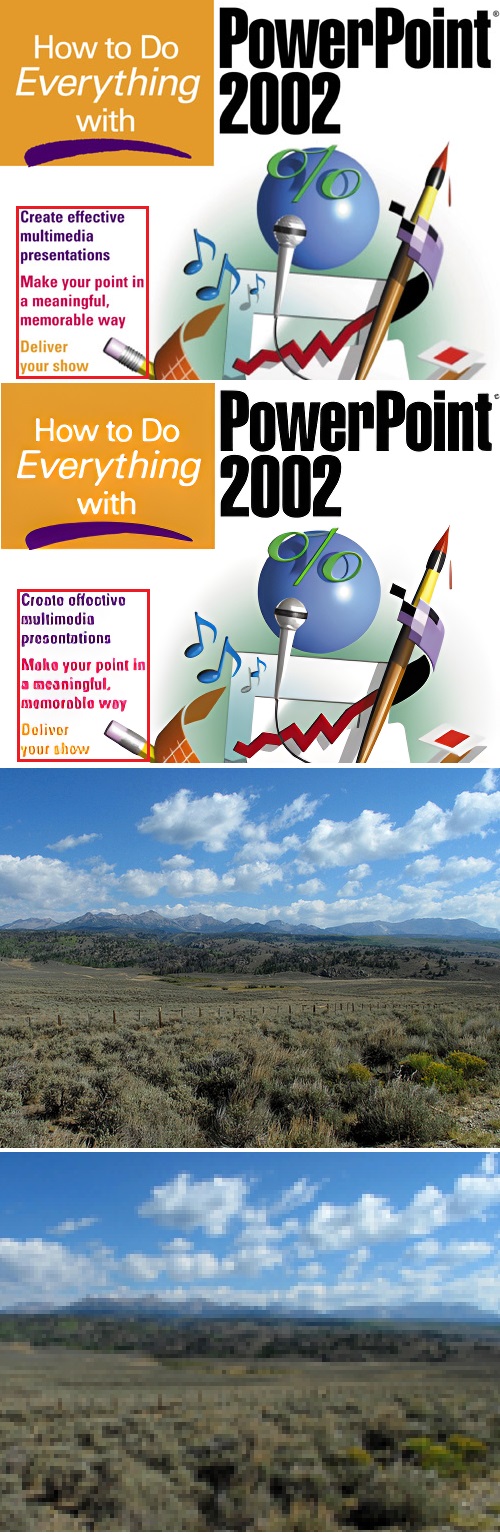} }\hspace{-1.5mm}
	\subfloat []{
		\includegraphics[height=4.85cm]{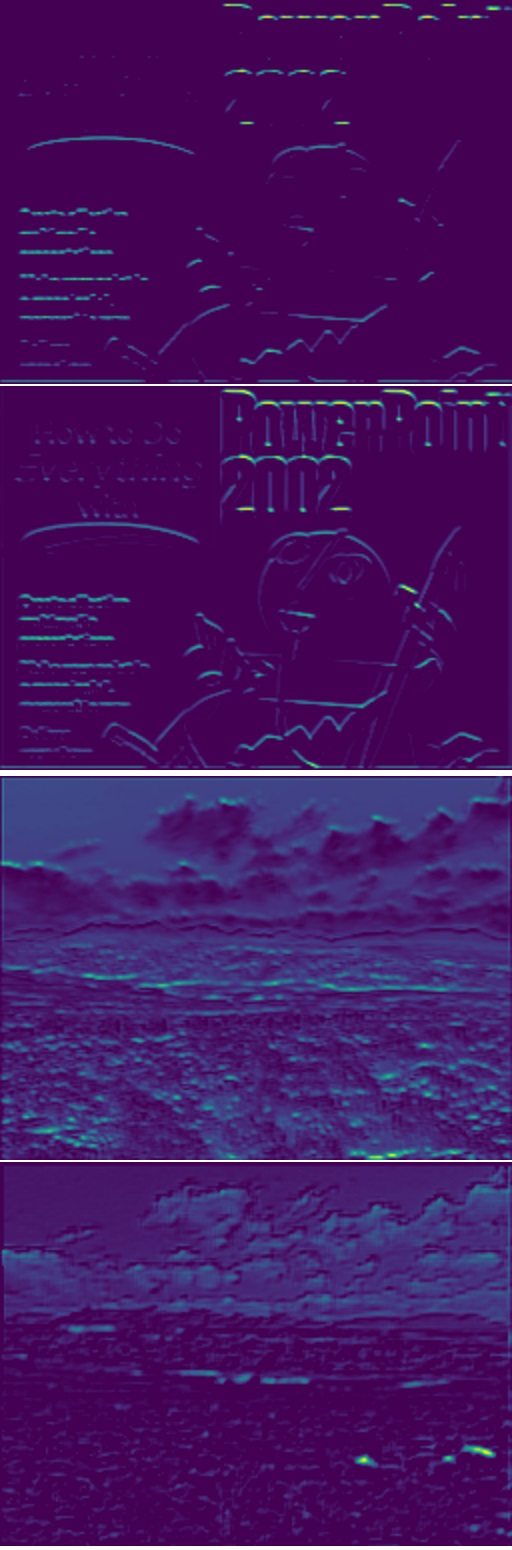} }\hspace{-1.5mm}
	\subfloat []{
		\includegraphics[height=4.85cm]{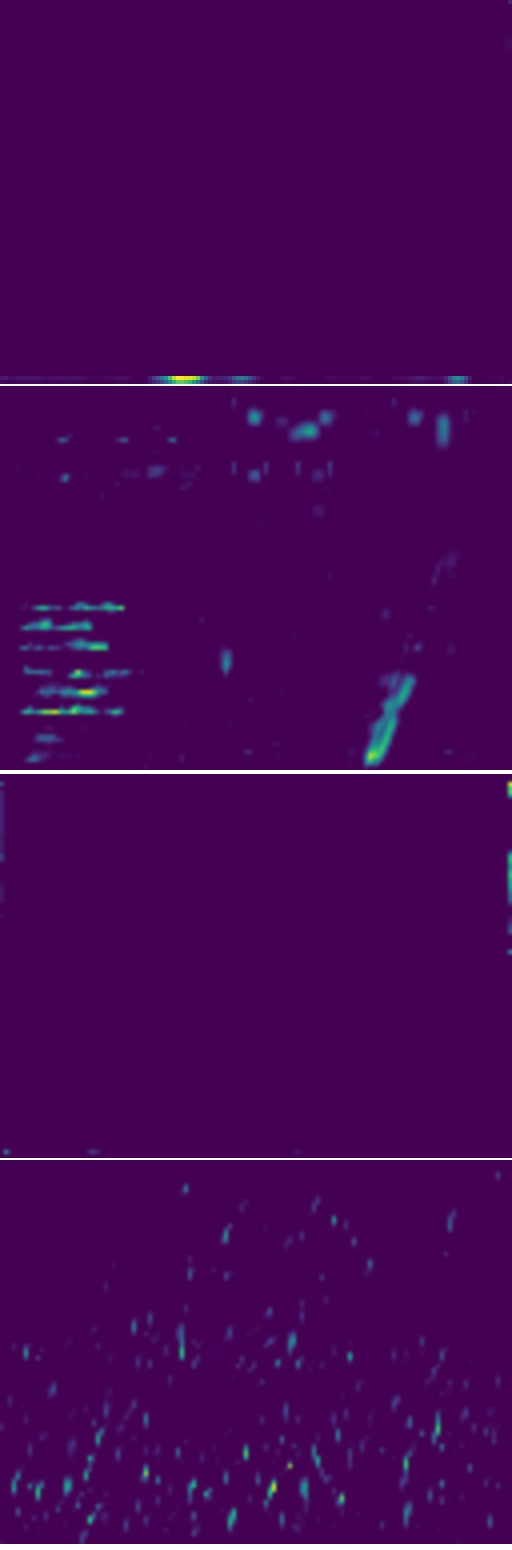} }\hspace{-1.5mm}
	\subfloat []{
		\includegraphics[height=4.85cm]{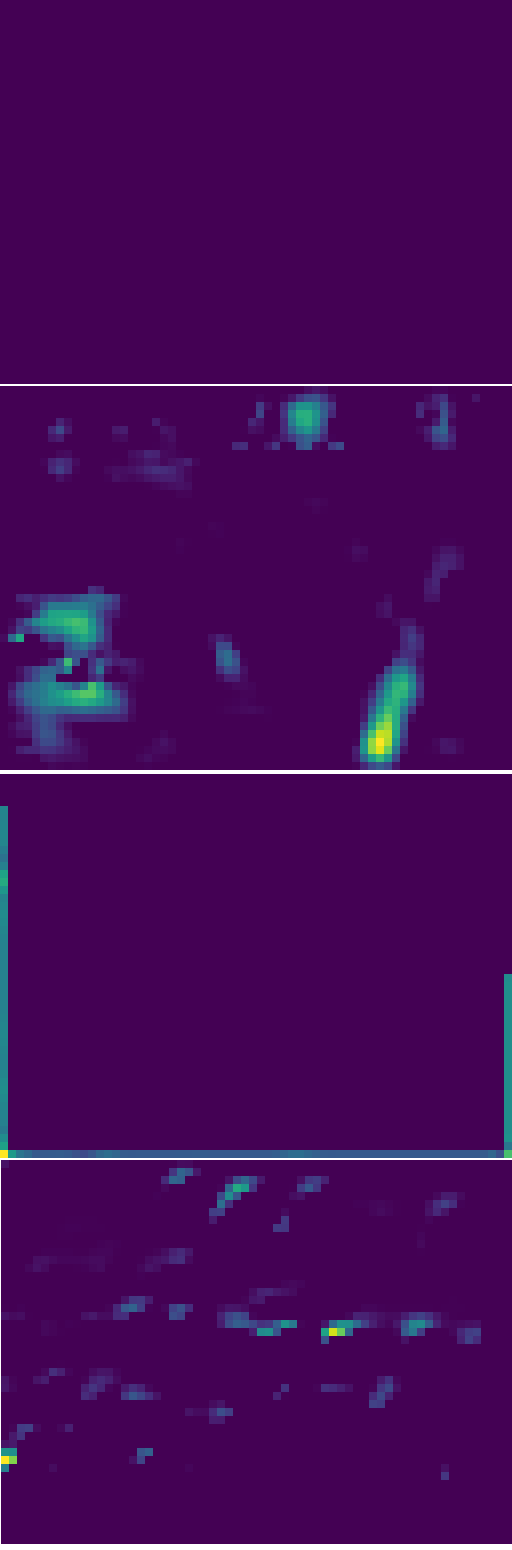} }\hspace{-1.5mm}
	\subfloat []{
		\includegraphics[height=4.85cm]{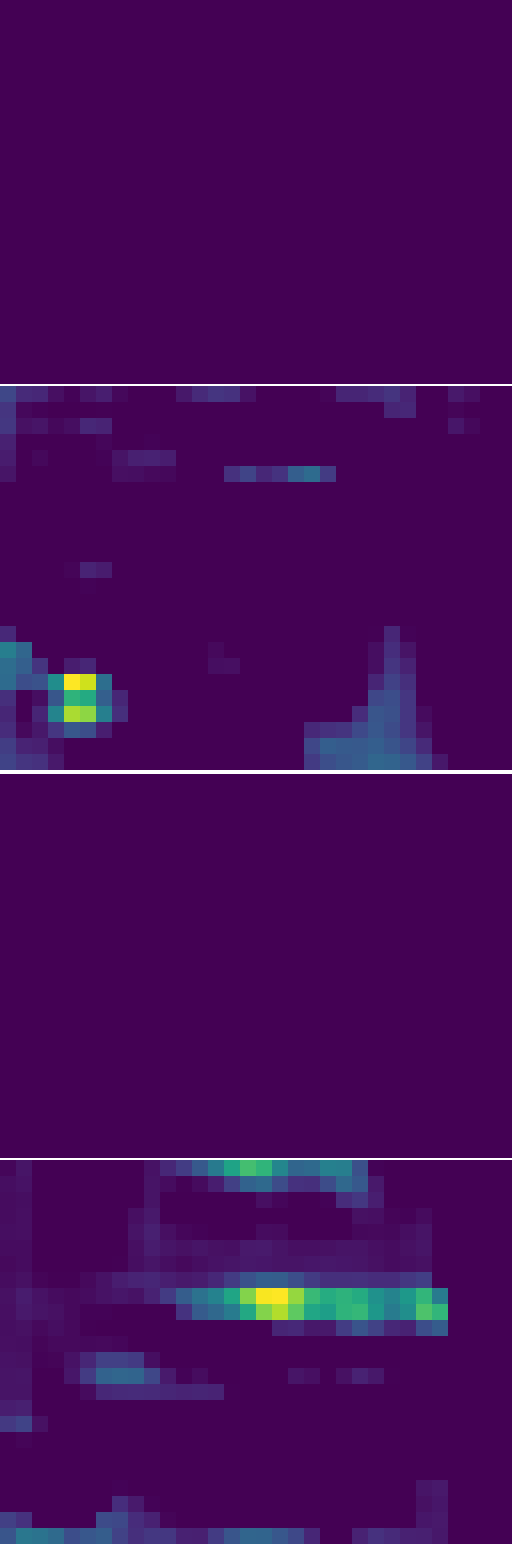} }\\
	\vspace{-0.3cm}
	\caption{Examples for multi-scale feature visualization of the structure score regressor. The odd and even rows of column (a) represent original HR images and SR images, respectively. The columns (b)-(e) are the multi-scale feature maps, and all images are zoomed in for visibility.}
	\label{fig:3}
\end{figure}

To intuitively show the captured feature, we present examples for multi-scale feature visualization of $M_s$ in Fig. \ref{fig:3}. The feature maps are sampled from the last convolutional layers of $i$th ($i=2,3,4,5$) stages of $F_r^i$ and $F_s^i$, respectively. From the figure, the lower features (Fig. \ref{fig:3} (b)) focus more on the structure information, while the higher features are more correlated to artifacts. Regarding SR images, the first SR image is clear but with obvious local structural difference compared with the reference image, and the second SR image has severe blocking artifacts. The above two representative examples generally occur in deep-learning-based and traditional SR methods, respectively. Encouragingly, the $M_s$ can effectively capture the artifacts (see Fig. \ref{fig:3} (c)-(e)). As a result, the design of $M_s$ is desirable.  

Considering that the features extracted from multi-scale have different expressions (see Fig. \ref{fig:3}), we first regress the scores for each scale separately, and then generate the final structure score by fusing the five scores. Deep IQA methods usually employ Multi-Layer Perception (MLP) to regress the image features to quality score. In this work, the MLP consists of three fully connected layers. The input of MLP should be a vector, so the feature maps $F_i$ should be converted to feature vector $F_{v,s}^i$. We empirically adopt the Global Average Pooling (GAP) to perform this conversion. The calculation process of structure score $S_s$ can be formulated as:      
\begin{equation}
\begin{aligned}
& F_{v,s}^i\gets {\rm{GAP}}_i(F_i),i\in \{1,2,3,4,5\},\\
& S_s^i={\rm{MLP}}_{s,i}(F_{v,s}^i;\theta_{s,mlp}^i),i\in \{1,2,3,4,5\},\\
& S_s=\sum_{i=1}^5(w_s^i\cdot S_s^i), i\in \{1,2,3,4,5\},
\end{aligned}
\label{func:4}
\end{equation}
where $\theta_{s,mlp}^i$ denotes the parameters of the ${\rm{MLP}}_{s,i}$. $w_s^i$ represents the weight of each scale score. From Eq. (\ref{func:4}), for each scale, we first convert the fused feature maps $F_{i}$ to a vector, then regress the vector to a structure score $S_s^i$. Obviously, $S_s^i$ and $F_{i}$ are highly correlated. From the Eq. (\ref{func:3}), $F_{i}$ is obtained by $F_r^i$ and $F_s^i$. Besides, we have reached a conclusion that the lower features focus more on structure information (see Fig. \ref{fig:3}). Thus, the network tends to assign higher weights to larger scales.  

\textbf{Perception score regressor}. It focuses on computing the perception score $S_p$ only for SR image $I_s$, and can be depicted as:
\begin{equation}
\begin{aligned}
S_p=M_p(I_s;\theta_p),
\end{aligned}
\label{func:6}
\end{equation}
where $M_p$ and $\theta_p$ represent the perception score regressor and its parameters. Considering that the HVS can automatically select the most salient or interested regions from natural scenes, employing visual saliency information of images is beneficial for quality assessment. Thus we develop a saliency-aided quality regressor $M_p$ using CNN. In particular, the $M_p$ shares the SR image ($I_s$) input branch with $M_s$. To avoid confusion, the $F_s^5$ of Eq. (\ref{func:2}), which is artifacts-aware for SR images (as shown in Fig. \ref{fig:3}), is denoted as $f_s$ and utilized as basic feature of $M_p$. Then, this paper resorts to the state-of-the-art DINet \cite{DINet} to generate a saliency map $I_{sal}$ for SR image. Subsequently, we fuse the saliency information to $M_p$ using element-wise multiplication between $I_{sal}$ and $f_s$. From Fig. \ref{fig:2}, the fused feature $F_{p}$ is activated by the saliency information, which is beneficial to our task of perceptual quality prediction. The calculation process of perception score $S_{p}$ can be formulated as: 
\begin{equation}
\begin{aligned}
&F_p \gets I_{sal} \otimes f_s,\\
&F_{v,p} \gets {\rm{GAP}} (F_p),\\
&S_p={\rm{MLP}}_p(F_{v,p};\theta_{p,mlp}),
\end{aligned}
\label{func:7}
\end{equation}
where $F_{p}$ indicates the fused features of $M_p$. $F_{v,p}$ denotes the converted vector from feature maps $F_{p}$, and $\theta_{p,mlp}$ represents the parameters of the ${\rm{MLP}}_p$.

\subsection{Adaptive Weight Regressor}
The adaptive weight regressor calculates adaptive weights to balance the structure score $S_s$ and perception score $S_p$. As discussed in Section 3, the adaptive weight regressor should be artifacts-aware. To meet the end-to-end requirements of the SPQE metric, we do not design an artifacts-aware weight regressor independent of SPQE. Instead, we resort to the features captured by $M_{p}$, which are exclusive to SR image itself. Rethinking the Fig. \ref{fig:3}, the even rows represent SR images and the feature maps extracted from the SR input branch of $M_{p}$/$M_s$ (they share the SR branch). The SR image \emph{powperpoint} is clear but with obvious local structural deformation, and the SR image \emph{grass} has severe blocking artifacts. These artifacts can be effectively captured by our model (see the even rows of Fig. \ref{fig:3} (c)-(e)). Thus, the artifacts-aware features $f_s$ of $M_p$ is adopted as the ultimate feature of the adaptive weight regressor $M_w$. Similar to $M_s$ and $M_p$, the perception weight $W_p$ can be obtained as:
\begin{equation}
\begin{aligned}
&F_{v,w} \gets {\rm{GAP}}(f_{s}),\\
&W_{p}={\rm{MLP}}_w(F_{v,w};\theta_{w,mlp}),
\end{aligned}
\label{func:9}
\end{equation}
where $F_{v,w}$ is the converted vector from $f_{s}$, and $\theta_{w,mlp}$ represents the parameters of the ${\rm{MLP}}_w$. Subsequently, the structure weight $W_s$ is computed as $1-W_p$, since the structure and perception scores serve as complementary one to the other in our SPQE metric.

\begin{figure}[t]
	\centering
	\subfloat []{
		\includegraphics[width=4.1cm]{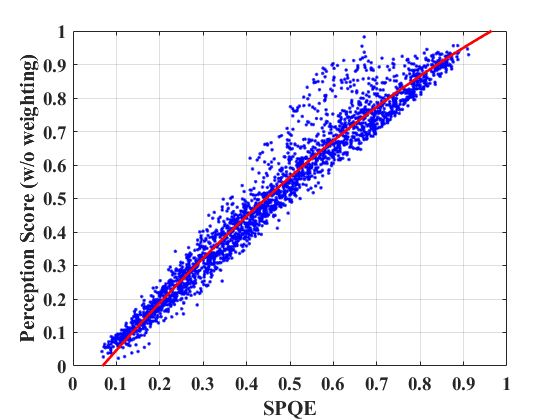} }\hspace{-2mm}
	\subfloat []{
		\includegraphics[width=4.1cm]{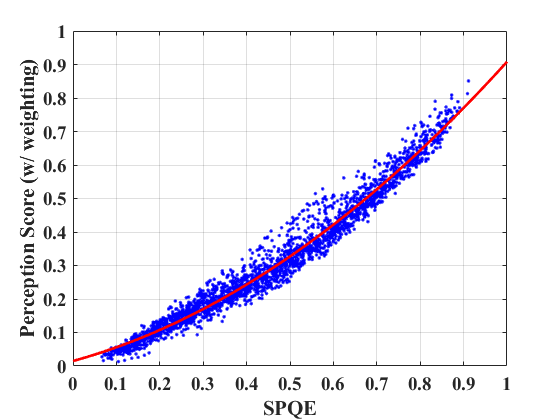} }\\
	\vspace{-0.3cm}
	\caption{Scatter plots of SPQE \emph{vs.} perception score.}
	\label{fig:5}
\end{figure} 

\begin{figure}[t]
	\centering
	\subfloat [0.7261/\textbf{\emph{0.9197}}/\textbf{0.8931}]{
		\includegraphics[width=2.6cm]{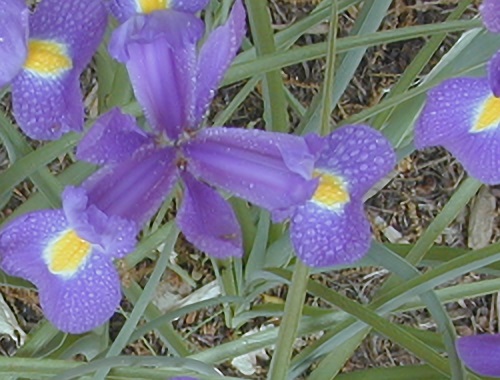} }\hspace{-1mm}
	\subfloat [0.6581/\textbf{\emph{0.6567}}/\textbf{0.6672}]{
		\includegraphics[width=2.6cm]{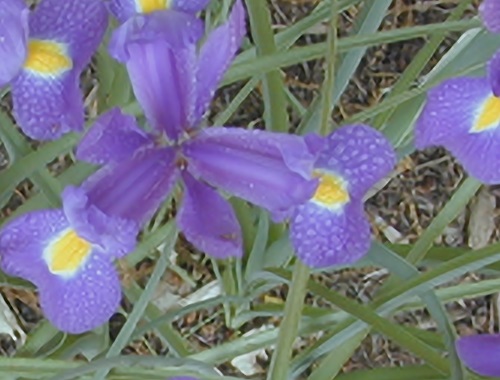} }\hspace{-1mm}
	\subfloat [0.7032/\textbf{\emph{0.1571}}/\textbf{0.1493}]{
		\includegraphics[width=2.6cm]{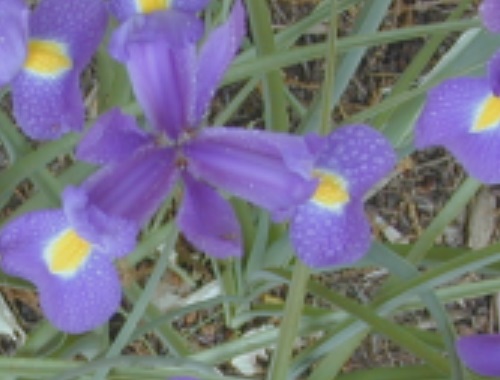} } \\
	\vspace{-2mm}
	\subfloat [0.7297/\textbf{\emph{0.9010}}/\textbf{0.9003}]{
		\includegraphics[width=2.6cm]{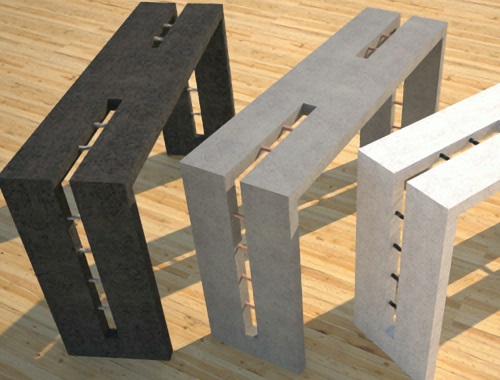} }\hspace{-1mm}
	\subfloat [0.6190/\textbf{\emph{0.6110}}/\textbf{0.6102}]{
		\includegraphics[width=2.6cm]{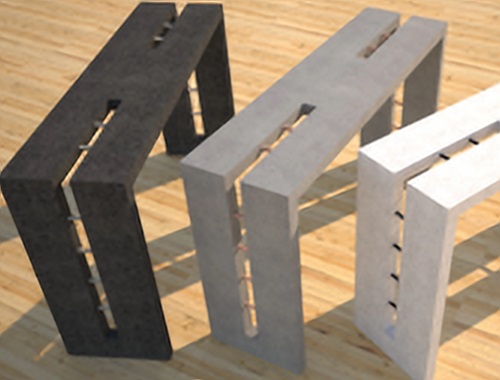} }\hspace{-1mm}
	\subfloat [0.6524/\textbf{\emph{0.2500}}/\textbf{0.2378}]{
		\includegraphics[width=2.6cm]{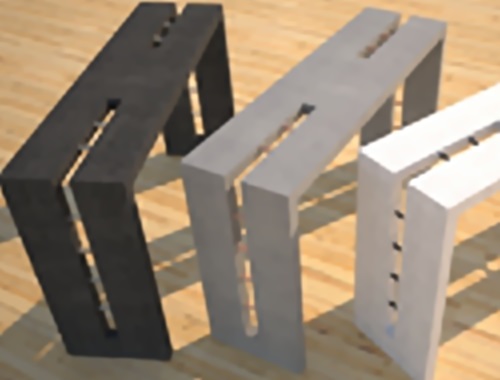} } \\
	\vspace{-0.3cm}
	\caption{Examples of perception weights on different SR images. The values of SR images represent perception weight, Mean Opinion Scores (MOS), and SPQE score, respectively.}
	\label{fig:6}
\end{figure}

\begin{table*}[t]
	\centering
	\caption{Performance comparison on four benchmark datasets}
	\label{table:2}
	\begin{tabular}{@{}clcc|cc|cc|cc|cc@{}}
		\toprule
		&	& \multicolumn{2}{c}{WIND}	& \multicolumn{2}{c}{CVIU}	&\multicolumn{2}{c}{QADS}&\multicolumn{2}{c}{SISAR}&\multicolumn{2}{c}{Average}     \\ \cmidrule(l){3-12} 
		\multirow{-2}{*}{Reference}&\multirow{-2}{*}{Methods} & PLCC & SRCC  & PLCC & SRCC	& PLCC & SRCC& PLCC & SRCC   & PLCC & SRCC  \\ \midrule
		HR&PSNR & 0.7906 & 0.7961& 0.5161 & 0.4783 & 0.3990 & 0.4119 & 0.6177&0.6112&0.5809&0.5744  \\
		HR&SSIM & 0.8171 & 0.7760& 0.5019 & 0.4289 & 0.4912 & 0.4773 & 0.5563&0.5422&0.5916&0.5561  \\
		HR&MS-SSIM &0.9332 & 0.9222& 0.7987&0.7861 &0.7515 &0.7476&0.6600&0.6436&0.7859&0.7749\\
		HR&FSIM & 0.9352&0.9142&0.7212&0.6981&0.7513&0.7476&0.7024&0.6901&0.7775&0.7625\\
		HR&GS  &0.8846&0.8608&0.6339&0.5974&0.6238&0.6162&0.6967&0.6825&0.7098&0.6892\\
		HR&GMSD &0.9003&0.8757&0.8437&0.8378&0.7882&0.7772&0.6957&0.6863&0.8070&0.7943\\
		$\times$&DIIVINE	&0.5814&0.5442&0.6193&0.5893&0.4588&0.4731&0.6134&0.6195&0.5682&0.5565\\
		$\times$&BRISQUE &0.5397&0.5085&0.6745&0.6552&0.5568&0.5851&0.5719&0.5603&0.5857&0.5773\\
		$\times$&BLINDS-II &0.5740&0.4984&0.4893&0.4572&0.3507&0.3290&0.5636&0.5868&0.4944&0.4679\\
		$\times$&NIQE &0.4565 &0.3311&0.5037&0.4791&0.1878&0.1122&0.6099&0.5735&0.4395&0.3740\\
		\midrule
		$\times$&LNQM    & - & -&{\textcolor{blue}{0.9651}}&{\textcolor{blue}{0.9633}}& - & -	& -&-&-& -\\
		$\times$&CNNSR &0.8217&0.7960&0.6628&0.6589&0.7183&0.7092&0.8040&0.7907&0.7517&0.7387\\
		$\times$&DeepSRQ &0.9566&{\textcolor{blue}{0.9223}}&0.8266&0.8071&0.8019&0.7959&0.6862&0.6436&0.8178&0.7922\\
		LR&NSS-SR  & 0.7733&0.6000&-&-&0.3670&0.2160&0.4507&0.3824&0.5303&0.3995\\
		LR&HYQM &0.6501&0.5635&-&-&0.3169&0.4341&0.4614&0.5145&0.5242&0.5040\\
		LR&DISQ &0.4569&0.3412& - & - &0.8501&0.8396&0.7136&0.7074&0.6735&0.6294	\\
		HR&SFSN &0.9429&0.8866&0.8638&0.8518&0.8308&0.8278&0.6635&0.6450&0.8253&0.8028\\
		LR&{\textbf{SPQE}}&{\textcolor{blue}{0.9590}}&0.9112&-&-&{\textcolor{blue}{0.9505}}&{\textcolor{blue}{0.9502}}&{\textcolor{blue}{0.9280}}&{\textcolor{blue}{0.9273}}&{\textcolor{blue}{0.9458}}&{\textcolor{blue}{0.9296}}\\
		HR&\textbf{SPQE}& {\textcolor{red}{0.9641}} & {\textcolor{red}{0.9317}} &{\textcolor{red}{0.9782}}&{\textcolor{red}{0.9776}}&{\textcolor{red}{0.9654}}&{\textcolor{red}{0.9625}}&{\textcolor{red}{0.9356}}&{\textcolor{red}{0.9363}}&{\textcolor{red}{0.9608}}&{\textcolor{red}{0.9520}}\\
		
		\bottomrule
	\end{tabular}
\end{table*}
To validate the effectiveness of the proposed adaptive tradeoff mechanism, we first show the scatter plots of $S_{spqe}$ \emph{vs.} $S_{p}$ on the test dataset of SISAR in Fig. \ref{fig:5}, then present examples of $W_{p}$ on different SR images in Fig. \ref{fig:6}. From Fig. \ref{fig:5}, the points of $S_{p}$ with weighting (Fig. \ref{fig:5} (b)) are more tightly distributed on the regression curve than the points of $S_{p}$ without weighting. It indicates that the $S_{p}$ with weighting are more correlated with the $S_{spqe}$. In Fig. \ref{fig:6}, the values of SR images indicate $W_{p}$/\textbf{\emph{MOS}}/$\bm{S_{spqe}}$, respectively. From the figure, images at the two ends of the quality range (Fig. \ref{fig:6} (a) (c) (d) (f)) have higher $W_{p}$ than images at the middle quality range (Fig. \ref{fig:6} (b) (e)). This confirms the three HVS characteristics discussed in Section 3. Besides, the $S_{spqe}$ values are very close to MOS. As indicated above, the adaptive tradeoff mechanism designed in this work is effective. 
\section{EXPERIMENTAL RESULTS}

\subsection{Experimental Setups}
\textbf{IQA datasets}. We perform experiments on four benchmark SR-IQA datasets, including WIND \cite{WIND} (312 SR images), CVIU \cite{CVIU} (1620 SR images), QADS \cite{QADS} (980 SR images) and SISAR \cite{SISAR} (12600 SR images). 

\textbf{Evaluation criteria}. To evaluate the performance of all IQA methods, we use two common criteria, including Pearson Linear Correlation Coefficient (PLCC) and SRCC, which are employed to measure the prediction accuracy and prediction monotonicity, respectively.

\textbf{Implementation details}. In each dataset, we randomly split the SR images into 80\% for training and 20\% for testing. Our SPQE metric is implemented with Keras, and trained on an NVIDIA GeForce RTX 3090 GPU. The training stage adopts the Adam optimizer \cite{kingma2014adam}, considering its prominent ability in handle sparse gradients and non-stationary objectives. Given the limited GPU memory, the batch size varies in accordance with the resolutions of input images. The initial learning rate is 0.0001, which is divided by 10 after every five epochs in the case that the validation loss does not decrease. The early stopping scheme is employed to avoid overfitting, and the \emph{patience} of epoch is set to 30. 

The reference images utilized in SPQE can be the original HR or LR images. In general, the original HR images exist only in laboratory environment. In practice, LR images can be adopted to provide partial reference information. The LR images should be upsampled first to enable the subsequent feature fusion process in $M_s$. The mainstream upsampling methods in image SR filed contain interpolation-based methods and learning-based methods \cite{SRsurvey}. The interpolation-based methods operate in image aspect, and are easy to implement. However, those methods are essentially a category of SR methods (see Section 2). The learning-based methods operate in the feature aspect, including deconvolution and sub-pixel convolution. The latter is usually adopted in the post-upsampling SR methods, since it needs multiple convolution operations in front of it. Thus, the deconvolution is selected in this work for upsampling the LR images. In practice, a deconvolution layer is added on the top of $M_{s}$. 

\begin{figure*}[t]
	\centering
	\subfloat [Original image]{
		\includegraphics[height=2.1cm]{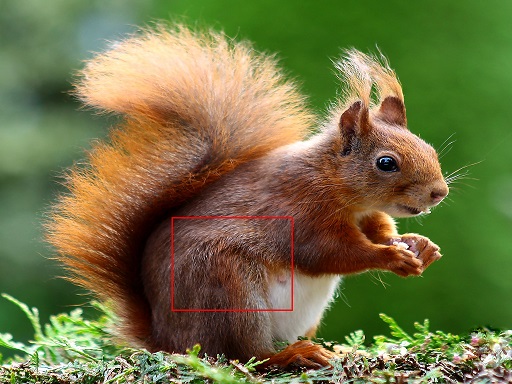} }\hspace{-1.5mm}
	\subfloat [Original]{
		\includegraphics[height=2.1cm]{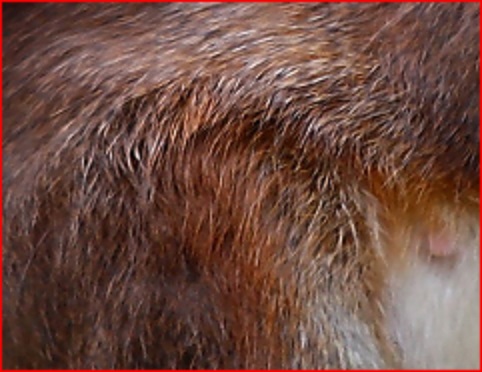} }\hspace{-1.5mm}
	\subfloat [0.8704/\textbf{0.3925}/\textcolor{red}{0.4110}]{
		\includegraphics[height=2.1cm]{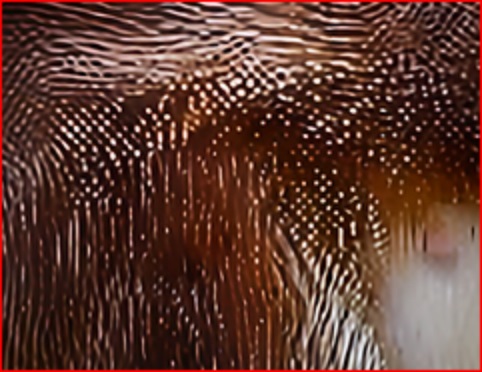} }\hspace{-1.5mm}
	\subfloat [0.8587/\textbf{0.5126}/\textcolor{red}{0.5299}]{
		\includegraphics[height=2.1cm]{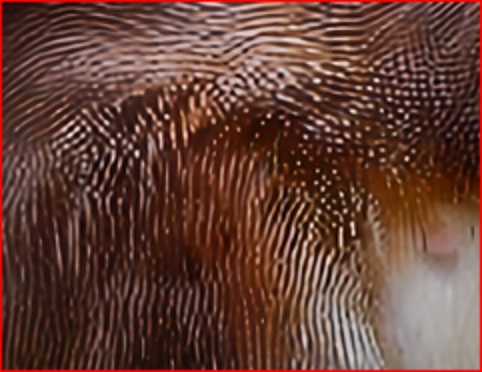} }\hspace{-1.5mm}
	\subfloat [0.8629/\textbf{0.7425}/\textcolor{red}{0.7677}]{
		\includegraphics[height=2.1cm]{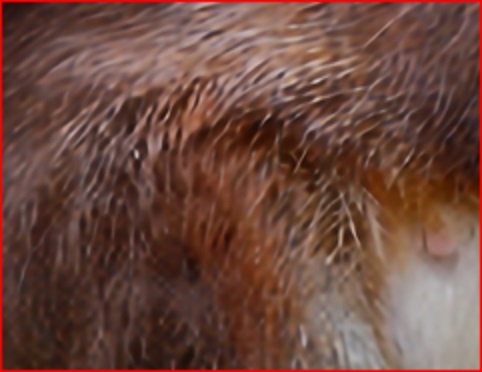} } \hspace{-1.5mm}
	\subfloat [0.8576/\textbf{0.7592}/\textcolor{red}{0.7758}]{
		\includegraphics[height=2.1cm]{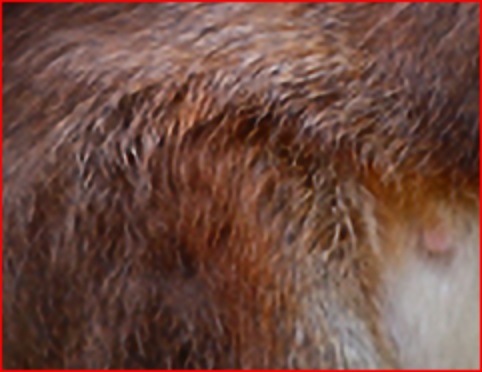} } \\ 
	\vspace{-2mm}
	
	\subfloat [Original image]{
		\includegraphics[height=2.1cm]{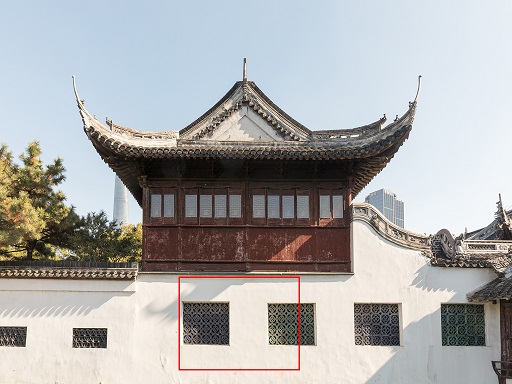} }\hspace{-1.5mm}
	\subfloat [Original]{
		\includegraphics[height=2.1cm]{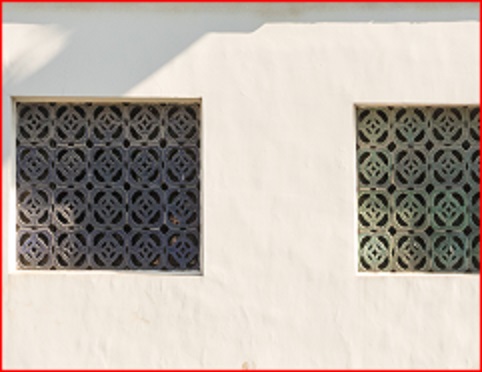} }\hspace{-1.5mm}
	\subfloat [0.8290/\textbf{0.5174}/\textcolor{red}{0.5285}]{
		\includegraphics[height=2.1cm]{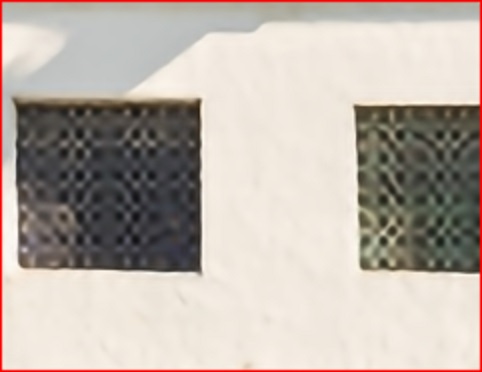} }\hspace{-1.5mm}
	\subfloat [0.8291/\textbf{0.7106}/\textcolor{red}{0.7270}]{
		\includegraphics[height=2.1cm]{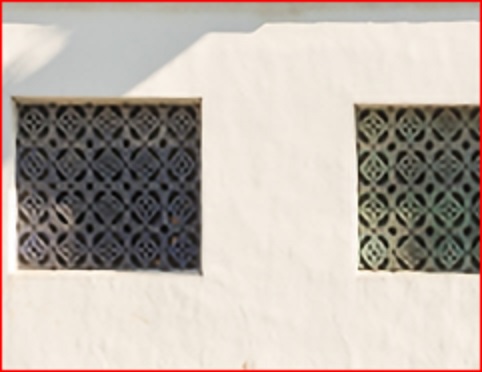} }\hspace{-1.5mm}
	\subfloat [0.8306/\textbf{0.7856}/\textcolor{red}{0.7807}]{
		\includegraphics[height=2.1cm]{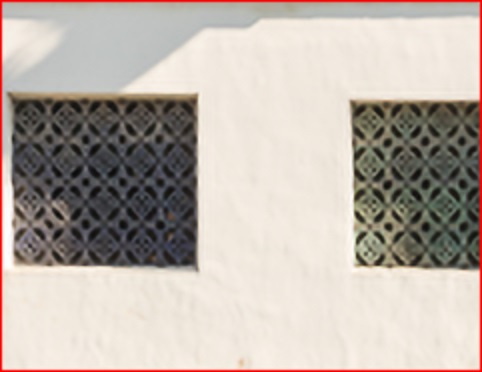} }\hspace{-1.5mm}
	\subfloat [0.8277/\textbf{0.8364}/\textcolor{red}{0.8345}]{
		\includegraphics[height=2.1cm]{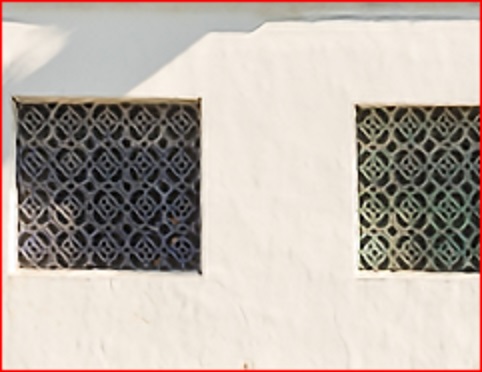} }\\
	\vspace{-2mm}
	
	\subfloat [Original image]{
		\includegraphics[height=2.1cm]{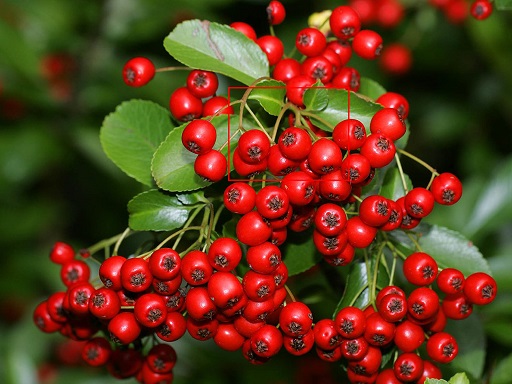} }\hspace{-1.5mm}
	\subfloat [Original]{
		\includegraphics[height=2.1cm]{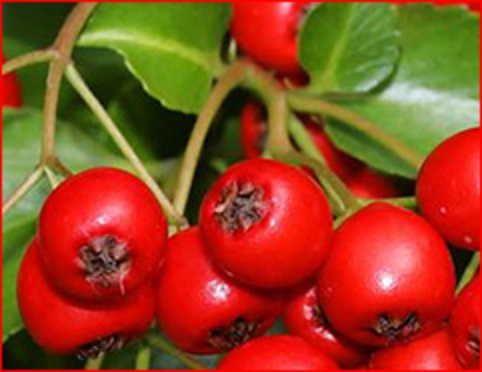} }\hspace{-1.5mm}
	\subfloat [0.7987/\textbf{0.4726}/\textcolor{red}{0.4712}]{
		\includegraphics[height=2.1cm]{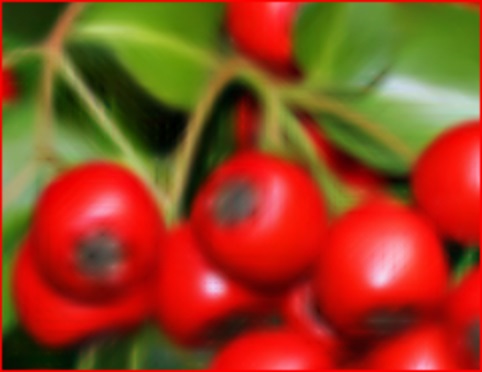} }\hspace{-1.5mm}
	\subfloat [0.8657/\textbf{0.6115}/\textcolor{red}{0.6198}]{
		\includegraphics[height=2.1cm]{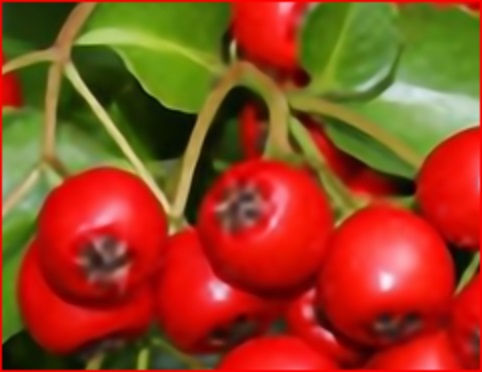} }\hspace{-1.5mm}
	\subfloat [0.8395/\textbf{0.7204}/\textcolor{red}{0.7518}]{
		\includegraphics[height=2.1cm]{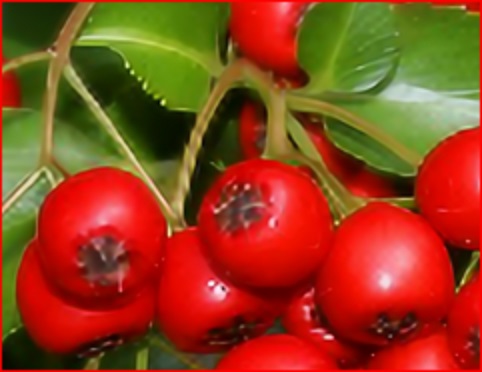} }\hspace{-1.5mm}
	\subfloat [0.8612/\textbf{0.8204}/\textcolor{red}{0.8460}]{
		\includegraphics[height=2.1cm]{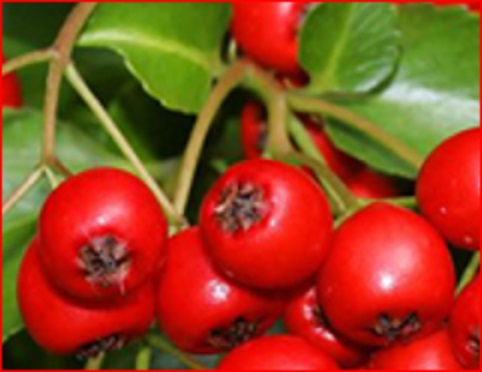} } \\ 
	\vspace{-2mm}
	
	\subfloat [Original image]{
		\includegraphics[height=2.1cm]{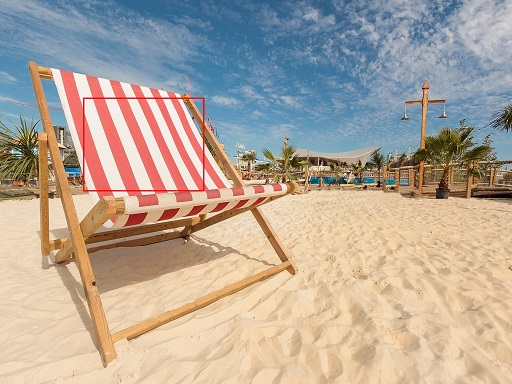} }\hspace{-1.5mm}
	\subfloat [Original]{
		\includegraphics[height=2.1cm]{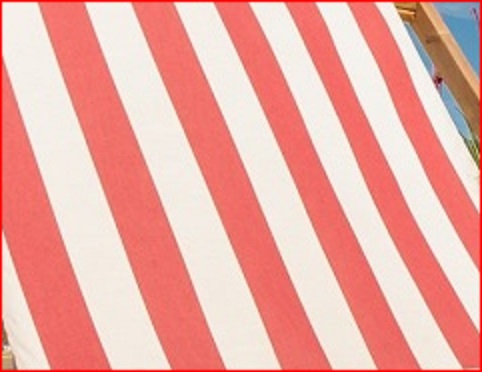} }\hspace{-1.5mm}
	\subfloat [0.7362/\textbf{0.4409}/\textcolor{red}{0.4490}]{
		\includegraphics[height=2.1cm]{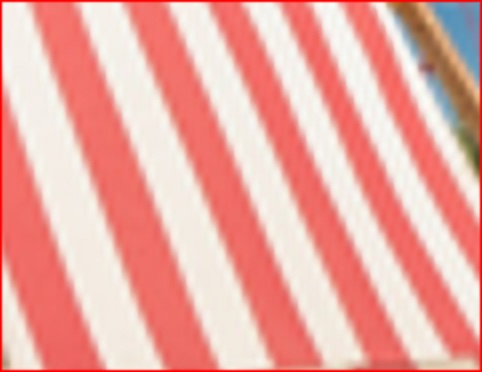}}\hspace{-1.5mm}
	\subfloat [0.7019/\textbf{0.5626}/\textcolor{red}{0.5936}]{
		\includegraphics[height=2.1cm]{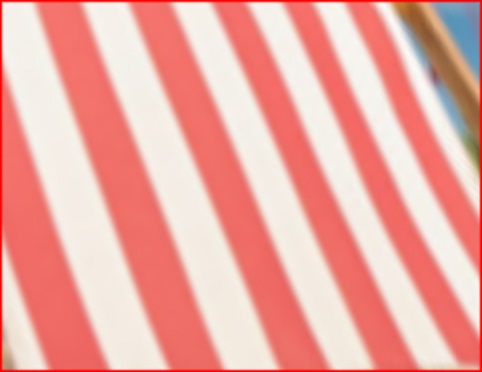}}\hspace{-1.5mm}
	\subfloat [0.7732/\textbf{0.7944}/\textcolor{red}{0.8024}]{
		\includegraphics[height=2.1cm]{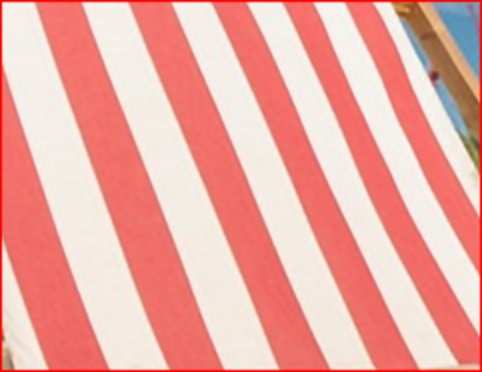}}\hspace{-1.5mm}
	\subfloat [0.7751/\textbf{0.8424}/\textcolor{red}{0.8468}]{
		\includegraphics[height=2.1cm]{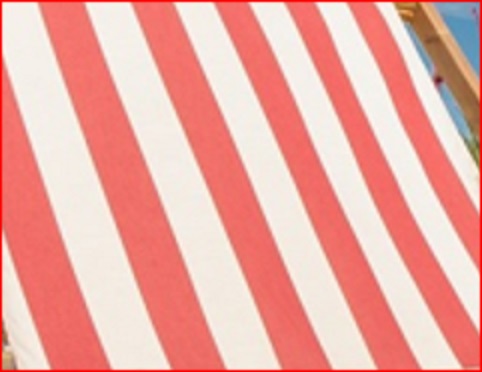} } \\ 
	
	\vspace{-3mm}
	\caption{Examples of predicted scores on the SISAR dataset: DISQ/SPQE (ours)/ground truth. The first column is original HR images, and all images are cropped for better visibility.}
	\label{fig:8}
\end{figure*} 
\subsection{Performance Comparison}
\textbf{Individual dataset evaluation}. Table \ref{table:2} shows the evaluation in individual dataset, where the best and 2nd-best results are labeled in \textcolor{red}{red} and \textcolor{blue}{blue}, respectively. The compared methods include six conventional FR-IQA (PSNR, SSIM \cite{SSIM}, MS-SSIM \cite{MS-SSIM}, FSIM \cite{FSIM}, GS \cite{GS}, GMSD \cite{GMSD}), four universal NR-IQA (DIIVINE \cite{DIIVINE}, BRISQUE \cite{BRISQUE}, BLIINDS-II \cite{BLIINDS-II}, NIQE \cite{NIQE}), and seven popular SR-IQA methods (NSS-SR \cite{WIND}, LNQM \cite{CVIU}, HYQM \cite{HYQM}, CNNSR \cite{CNNSR}, DeepSRQ \cite{DeepSRQ}, SFSN \cite{SFSN}, DISQ \cite{SISAR}). For fair comparison, all compared results are obtained by executing their public codes or citing from their papers. For the machine-learning based LNQM method, its training code is publicly unavailable, thus only results on its training set CVIU are provided. Note that in this paper, the utilized training and testing datasets of all learning-based methods are identical. In addition, the LR images do not exist in CVIU dataset, hence the results of four RR-IQA methods on CVIU are not available. From Table \ref{table:2}, we have the following findings.

\textbf{First}, the performance of SR-IQA metrics outperforms the ten conventional IQA methods. This is because the conventional IQA algorithms are designed for images that contain common artifacts, such as encoding distortion, \emph{etc}. However, the artifacts introduced by SR is complicated and mixed. As a result, the performance of those general-purpose IQA methods on SR images is not competitive. \textbf{Second}, the proposed SPQE metric is superior to the seven SR-IQA methods. Generally, these metrics do not consider the tradeoff between the no-reference and referenced scores when evaluating the quality of SR images. Thus the experimental results of these methods are not desirable. \textbf{Third}, the overall performance of the SPQE metric is still optimal and impressive. For the average SRCC, the SPQE metric with HR as reference outperforms the third best method by a large margin of 0.15. This indicates the superiority of our SPQE metric, which embraces the benefits of the elaborately designed quality regressors and adaptive tradeoff mechanism.

For visual comparison, we show the predicted scores of our SPQE metric compared to the DISQ model and ground truth in Fig. \ref{fig:8}. From the figure, the DISQ may prefer images with prominent texture. It gives high scores to the Fig. \ref{fig:8} (c) (e), which have obvious structural distortions. In contrast, the SPQE prefers Fig. \ref{fig:8} (l) (x). This is because the SPQE metric can adaptively balance the quality scores between the no-reference perceptual quality and the referenced structural similarity. Moreover, our SPQE metric has a lower prediction deviation compared with the DISQ model. The slight gap between the SPQE scores and ground truth indicates that our SPQE metric achieves superior performance in SR-IQA.

\textbf{Cross-dataset evaluation}. To verify the generalization ability of the proposed SPQE metric, we conduct cross-dataset evaluation experiments and compare the results of four data-driven SR-IQA algorithms. In this experiment, methods are trained on one dataset and validated on others. The SRCC results of the cross-dataset evaluation are shown in Table \ref{table:3}, where the best and 2nd-best results are labeled in \textcolor{red}{red} and \textcolor{blue}{blue}, respectively. From Table \ref{table:3}, the proposed SPQE metric shows preferable generalization ability on the four benchmark datasets, even when training on a small dataset (WIND/QADS) and testing on the large-scale SISAR dataset. It indicates the strong generalization ability of our SPQE metric in predicting the quality of unknown images. The generalization abilities of the compared methods are not competitive, this may be due to the simple structure design strategies in these deep-learning-based methods.
\begin{table*}[t]
	\centering
	\caption{SRCC results in the cross-dataset evaluation}
	\label{table:3}
	\begin{tabular}{@{}l|l|ccc|ccc|ccc|ccc@{}}
		\toprule
		&Train	& \multicolumn{3}{c|}{WIND}	& \multicolumn{3}{c|}{CVIU}	&\multicolumn{3}{c|}{QADS} &\multicolumn{3}{c}{SISAR}        \\ 
		\cmidrule(l){2-14} 
		\multirow{-2}{*}{Ref.}&Test	& CVIU&QADS&SISAR	&WIND&QADS&SISAR	&WIND&CVIU&SISAR &WIND&CVIU&QADS \\ \cmidrule(l){1-14} 
		$\times$&DeepSRQ &{\textcolor{blue}{0.5385}}&0.4586&0.4175&{\textcolor{blue}{0.7322}}&0.5158&0.4601&0.5836&{\textcolor{blue}{0.6481}}&0.2304&{\textcolor{blue}{0.7839}}&{\textcolor{blue}{0.6495}}&0.4666\\
		$\times$&CNNSR &0.3296&0.2646&0.1315&0.3697&0.2351&0.2364&0.1788&0.2126&0.1259&0.5168&0.5420&0.3637\\
		$\times$&LNQM &-&-&-&0.7010&{\textcolor{red}{0.7255}}&{\textcolor{red}{0.6658}}&-&-&-&-&-&-\\
		LR&DISQ &-&0.4793&0.4648&-&-&-&{\textcolor{blue}{0.7447}}&-&0.4482&0.7521&-&{\textcolor{blue}{0.7220}}\\
		LR&\textbf{SPQE}&-&{\textcolor{red}{0.5703}}&{\textcolor{blue}{0.5048}}&-&-&-&0.5954&-&{\textcolor{blue}{0.5332}}&0.6052&-&0.6117\\
		HR&\textbf{SPQE}& {\textcolor{red}{0.5577}}&{\textcolor{red}{0.7189}}&{\textcolor{red}{0.5924}}&{\textcolor{red}{0.7600}}&{\textcolor{blue}{0.6258}}&{\textcolor{blue}{0.6008}}&{\textcolor{red}{0.7594}}& {\textcolor{red}{0.6936}}&{\textcolor{red}{0.6085}}&{\textcolor{red}{0.8024}}&{\textcolor{red}{0.7548}}&{\textcolor{red}{0.7763}}\\	
		\bottomrule
	\end{tabular}
\end{table*}

\subsection{Ablation Study}
We conduct ablation experiments on the SISAR \cite{SISAR} dataset, given that the SISAR is the largest-ever SR-IQA dataset and appropriate for evaluating the performance of deep-learning-based methods. In this section, the SPQE metric adopts original HR images as reference by default.

\textbf{Contributions of quality score regressors}. We separately test the performance of the structure score regressor and perception score regressor in Table \ref{table:4}. The two regressors are trained under the same parameter settings. From Table \ref{table:4}, we can draw two conclusions. \textbf{Firstly}, the model achieves desirable performances even with one type of regressor. Compared with the results on the SISAR dataset in Table \ref{table:2}, the corresponding performance still outperforms other IQA methods in SR-IQA. \textbf{Secondly}, by adaptively balancing the no-reference perceptual quality and referenced structural similarity scores modeled by the two regressors, our SPQE metric achieves the best performance of all. This demonstrates the effectiveness of our overall framework design.

\textbf{Contributions of adaptive tradeoff mechanism}. We verify the effect of the weight setting method for perception and structure scores by comparing six different weight setting methods in Table \ref{table:5}, from which, we can draw two conclusions. \textbf{Firstly}, towards balancing the perception and structure quality scores of images, utilizing different ratios of $W_{p}$ and $W_{s}$ always achieves better performance than a simple average ($W_{p} \ 0.5 \ \& \ W_{s} \ 0.5$). \textbf{Secondly}, by employing the adaptive weight, our SPQE metric achieves the optimal performance of all. This fact indicates the superiority of the adaptive tradeoff mechanism used in the SPQE metric.
\begin{table}[t]
	\centering
	\caption{Performance on quality score regressor}
	\label{table:4}
	\begin{tabular}{@{}c|c|c|cc@{}}
		\toprule
		\multicolumn{2}{c|}{Quality Regressor}&&&\\
		\cmidrule(l){1-2}
		Structure&Perception&\multirow{-2}{*}{ \begin{tabular}[c]{@{}c@{}}Adaptive Tradeoff\\Mechanism\end{tabular}}&\multirow{-2}{*}{PLCC}&\multirow{-2}{*}{SRCC}\\	\midrule
		$\checkmark$ &&&0.8392&0.8372\\
		&$\checkmark$&&0.8895&0.8900\\
		$\checkmark$&$\checkmark$&$\checkmark$&\textcolor{red}{0.9356}&\textcolor{red}{0.9363}\\
		\bottomrule
	\end{tabular}
\end{table}
\begin{table}[t]
	\centering
	\caption{Performance on weight setting method}
	\label{table:5}
	\setlength{\tabcolsep}{5mm}
	\begin{tabular}{c|cc}
		\toprule
		Weight Setting & PLCC & SRCC   \\ \midrule
		$W_{p}$ 0.2 \ \& \ $W_{s}$ 0.8  &0.8717&0.8713 \\
		$W_{p}$ 0.4 \ \& \ $W_{s}$ 0.6  &0.9048&0.9053\\
		$W_{p}$ 0.5 \ \& \ $W_{s}$ 0.5  &0.8822&0.8829\\
		$W_{p}$ 0.6 \ \& \ $W_{s}$ 0.4  &0.8969&0.8974\\
		$W_{p}$ 0.8 \ \& \ $W_{s}$ 0.2  &0.8945&0.8947\\
		Adaptive Weight&\textcolor{red}{0.9356}&\textcolor{red}{0.9363}\\
		\bottomrule
	\end{tabular}
\end{table}

\textbf{Contributions of multi-scale and saliency strategies}. We also examine the impact of multi-scale and saliency strategies. The multi-scale strategy represents the feature extraction, feature fusion, and score regression methods of structure score regressor. The saliency strategy refers to that we leverage salieny maps to fuse perception features in perception score regressor. We compare four different strategy combination methods in Table \ref{table:6}, from which, we can draw two conclusions. \textbf{Firstly}, both the multi-scale and saliency information are beneficial in evaluating the quality of SR images. \textbf{Secondly}, by embedding the two strategies in structure and perception quality score regressors, our SPQE metric outperforms the other methods by a large margin. These findings demonstrate the superiority of the multi-scale and saliency strategies employed in our SPQE metric.
\begin{table}[t]
	\centering
	\caption{Performance on multi-scale and saliency strategies}
	\label{table:6}
	\setlength{\tabcolsep}{4mm}
	\begin{tabular}{@{}cc|cc|cc@{}}
		\toprule
		\multicolumn{2}{c|}{Multi-scale}&\multicolumn{2}{c|}{Saliency}&&\\
		\cmidrule(l){1-4}
		$w/$&$w/o$&$w/$&$w/o$&\multirow{-2}{*}{PLCC}&\multirow{-2}{*}{SRCC}\\	\midrule
		$\checkmark$ &&&$\checkmark$&0.8406&0.8409\\
		&$\checkmark$&$\checkmark$&&0.8325&0.8313\\
		&$\checkmark$&&$\checkmark$&0.8430&0.8418\\
		$\checkmark$&&$\checkmark$&&\textcolor{red}{0.9356}&\textcolor{red}{0.9363}\\
		\bottomrule
	\end{tabular}
\end{table}


\section{CONCLUSIONS}
Recent advances of image SR techniques call for an efficient SR-IQA. In this paper, we propose a unified, end-to-end metric named SPQE for SR-IQA. We give a theoretical analysis on how to balance the quality scores between no-reference perceptual quality and referenced structural similarity of SR images. To model the two types of quality scores, we design a perception score regressor and a structure score regreessor by leveraging saliency and multi-scale information, respectively. Further, we develop an adaptive weight regressor to calculate adaptive weights for the two types of quality scores. Extensive experimental results on four benchmark datasets have demonstrated the impressive performance of our SPQE metric, which is superior to the state-of-the-art SR-IQA methods.  

\balance
\bibliographystyle{ACM-Reference-Format}
\bibliography{ref1}
\end{document}